\documentclass[12pt]{article}
\usepackage{graphicx}
\usepackage{epsfig}
\newcommand{\ccdot}{ \! \cdot \!}

\begin{document}

\title{{\bf Form factors in RQM approaches:\\
 constraints from space-time translations}} 
\author{ 
B.  Desplanques$^{1}$\thanks{{\it E-mail address:}  desplanq@lpsc.in2p3.fr},
Y.B. Dong$^{2}$\thanks{{\it E-mail address:}  dongyb@mail.ihep.ac.cn}
\\
$^{1}$LPSC, Universit\'e Joseph Fourier Grenoble 1, CNRS/IN2P3, INPG,\\
  F-38026 Grenoble Cedex, France \\
$^{2}$Institute of High Energy Physics, Chinese Academy of Science,\\ 
Beijing 100049, P. R. China}

\sloppy

\maketitle

\begin{abstract}
\small{
\noindent
Different relativistic quantum mechanics approaches have recently 
been used to calculate properties of various systems, form factors 
in particular. It is known that predictions, which most often rely on
a single-particle current approximation, can lead to predictions with 
a very large range. It was shown that accounting for constraints related 
to space-time translations could considerably reduce this range. 
It is shown here that predictions can be made identical for a large range 
of cases. These ones include the following approaches: instant form,  
front form, and  ``point-form" in arbitrary momentum configurations 
and a dispersion-relation approach which can be considered 
as the approach which the other ones should converge to. 
This important result supposes both an implementation 
of the above constraints and an appropriate single-particle-like current. 
The change of variables that allows one to establish the equivalence 
of the approaches is given. Some points are illustrated with numerical 
results for the ground state of a system consisting of scalar particles. }
\end{abstract} 
\section{Introduction}
It is known that the different implementations of relativity 
which are currently under discussion should ultimately lead to identical 
predictions for properties of various systems, such as form factors 
for hadronic ones.
In the case of relativistic quantum mechanics (RQM), which we are 
concerned with here, many forms were proposed \cite{Dirac:1949cp} 
and discussed~\cite{Keister:sb}. 
Looking at their predictions for form factors, it is observed 
that the above convergence is far to be achieved for calculations based 
on a one-body-like current. This has been shown for instance in the framework 
of a theoretical model \cite{Amghar:2002jx}, using the same solution of a mass
operator. In the case of the pion charge form factor, reasonable wave functions
describing measurements could be found in the front and instant forms 
but none in the ``point form"~\cite{Junhe:2004}. 
A somewhat similar conclusion holds for the nucleon form factors where 
measurements could be described with a spatially extended wave function 
in the instant and front forms whereas a compact one was required 
in the ``point form"~\cite{Julia-diaz:2004}. Confirming this result 
in some sense, it was shown that estimates of nucleon form factors 
using the same wave function in the instant form and in the ``point-form" 
were quite different \cite{Plessas:2004}.
For the deuteron case, which is less relativistic 
than the previous systems, a conclusion supposes to disentangle the role 
of dynamics, nucleon form factors and relativistic approaches. 
One can nevertheless guess that the drop off of form factors 
at very high momentum transfer in  the ``point-form" \cite{Allen:2000ge} 
is faster than in other forms \cite{Chung:1988,Vanorden:1995}, 
emphasizing an effect mentioned in the first reference.
The lack of convergence in all cases indicates that many-body currents 
should play a si\-gni\-ficant and sometimes essential role. Interes\-ting\-ly, 
Lorentz invariance of form factors does not necessarily imply
good results. Instead, it was noticed that the discrepancies could stem from 
a violation of constraints related to space-time translations,
going beyond energy-momentum conservation which is generally 
assumed~\cite{Desplanques:2004sp}.  
Implementing these constraints which are rarely considered 
could largely remove discrepancies between different approaches. 
An important question is whether they can be 
completely removed  and what has to be done in this order. 

Besides the instant-, front- and point-form approaches, 
a dispersion-relation based approach has been considered in the literature, 
see for instance works  by Anisovich  {\it et al.} \cite{Anisovich:1992}  
or by Krutov and Troitsky \cite{Krutov:2002} and references therein. 
At some point, this approach should evidence some relationship  
with RQM approaches. In this respect, the last work is probably closer 
to our motivations. It was considered as an instant-form one 
by the authors but it was shown independently by Melikhov \cite{Melikhov} 
and one of the present author (B.D.) that form factors obtained 
in this approach were identical 
to the standard front-form ones (with $q^+=0$), using an appropriate 
change of variables. Moreover, the absence of reference to a particular
4-dimensional orientation of some  hyperplane, which underlies the instant 
and  front forms, could suggest that the approach was rather 
of the Dirac point-form type (hyperboloid surface). Actually, 
in the above dispersion-relation approach, the interaction 
with the external field is described by a current involving 
a system made of free particles. This essential ingredient 
is independent of the interaction effects that are here or there in the
different RQM forms. It is also noticed that the formalism relies 
on on-mass shell particles and uses wave functions that can be identified 
to the solutions of a mass operator. It thus sounds that the  form factors 
in the dispersion-relation approach under consideration here 
are form independent and could be those which the other ones 
should converge to.

In the present paper, we look at the relationship of form factors calculated 
in different RQM forms and in the dispersion-relation approach, 
and show how the equivalence of these approaches can be achieved. 
This is illustrated  with the $J=0$ state of a system composed 
of scalar particles (ground state of the Wick-Cutkosky model 
\cite{Wick:1954,Cutkosky:1954}
for numerical considerations). Such a system is characterized 
by charge and scalar form factors, $F_1(Q^2)$ and $F_0(Q^2)$ 
respectively. The first one, which is closely related to the definition 
of the normalization, is considered in all forms. The second one, 
which can provide some complimentary information, is considered 
in a few cases only. 

The plan of the paper is as follows. In the second section, 
we remind the main features of the dispersion-relation approach 
that are relevant here and give the corresponding expression of form factors 
$F_1(Q^2)$ and $F_0(Q^2)$. Though the relation to this approach came 
at the latest stage of our work, we feel it is appropriate to begin 
with its presentation as it could be the convergence point of different
approaches. The third section is devoted to the constraints related 
to space-time translations in RQM approaches 
while their implementation in the calculation of form factors is discussed in
the fourth section. In the fifth section, we show 
how the identity of RQM results for form factors to dispersion-relation ones 
can be made in the Breit-frame case. The change of variables that is implied 
in this identity is made explicit for each RQM approach. 
Some points, like the sensitivity to the choice of the current 
or the dependence of the effects on the approach, 
are illustrated numerically in the sixth section for the Breit-frame case. 
The seventh section contains the conclusion. Many details 
and generalizations to arbitrary momentum configurations 
are considered in three appendices. Due to the novelty of the considerations
under discussion here, the absence of a common method for all cases and our
intent to limit equations to essential ones in the main text, 
some of the appendices are somewhat long.

\section{The dispersion-relation approach}

Dispersion relations are a powerful tool based on fundamental properties 
of the underlying theory. They often provide results that would require 
considerable effort in other approaches. This formalism is considered 
in this section for calculating the form factors of the $J=0$ state 
composed of two scalar constituents with mass $m$.
They are denoted $F_1(Q^2)$ and $F_0(Q^2)$ 
and we follow ref. \cite{Amghar:2002jx} for their definitions.
Concerning the application of dispersion relations 
to the calculation of the charge form factor, $F_1(Q^2)$, we refer 
to ref.~\cite{Krutov:2002} for details. It is noticed that the expression 
so obtained assumes current conservation and the simplest one-body 
current, $ <p_i|J^{\mu}|p_f> \propto (p_i+p_f)^{\mu}$. The
application of the approach to the scalar form factor, $F_0(Q^2)$, is much
simpler as complications relative to current conservation are absent. 

\begin{figure}[htb]
\begin{center}
\includegraphics[width=0.48\textwidth]{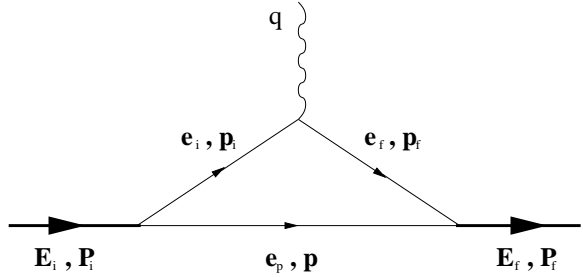}
\end{center}
\caption{Interaction  with an external field together with kinematic
notations.\label{fig1}}
\end{figure}

The expressions of form factors can be expressed in terms 
of the Mandelstam variables $s_i=(p_i\!+\!p)^2$, $s_f=(p_f\!+\!p)^2$, 
and $q^2=(p_i\!-\!p_f)^2\;(=-Q^2)$, where $p_i^{\mu},\;p_f^{\mu}$ 
represent the on-mass shell momenta of the constituents interacting 
with the external field and $p^{\mu}$ the momentum 
of the spectator constituent  (see fig. \ref{fig1} for notations). 
These relations are helpful in making the relation of the
dispersion-relation approach considered in this section and RQM approaches
discussed in the following sections. The expressions of form factors thus read:
\begin{eqnarray} 
F_1(Q^2) &\!=\!& \frac{1}{N} \int \!\!\int ds_i\;ds_f \; 
\frac{ \;Q^2\,\Big(s_i\!+\!s_f+Q^2\Big)\;\theta(\cdots) }{
\Big((s_i\!-\!s_f)^2+2Q^2\,(s_i\!+\!s_f)+Q^4 \Big)^{3/2} }\; 
\phi(s_f) \, \phi(s_i)\,, 
\label{eq:ff1-disp}
 \\
F_0(Q^2) &\!=\!& \frac{1}{N} \int \!\! \int  ds_i\;ds_f \;  
 \frac{\;\theta(\cdots)}{2\, 
\Big((s_i\!-\!s_f)^2+2Q^2\,(s_i\!+\!s_f)+Q^4 \Big)^{1/2} } \;
 \phi(s_f) \, \phi(s_i)  \, , 
\label{eq:ff0-disp}
\end{eqnarray}
where:
\begin{eqnarray} 
N &=& \int ds \;\sqrt{\frac{s-4\,m^2}{s}}\;\phi^2(s) \,,\label{eq:norm}
\\
\theta(\cdots)&=&\theta 
\Big(\frac{s_i\;s_f}{D} -m^2\Big) , \hspace*{0.5cm} 
 {\rm with}\;\; D=2(s_i\!+\!s_f)\!+\!Q^2\!+\!\frac{(s_i\!-\!s_f)^2}{Q^2}\, .
\label{eq:theta}
\end{eqnarray}
The overall writing of the above expressions is equivalent to that one 
given in ref. \cite{Krutov:2002} but the presentation retained here 
is more symmetrical and also involves some significant 
corrections\footnote{The expression of $F_1(Q^2)$ given in 
ref. \cite{Krutov:2002} contains an extra factor 
$(s_i\!+\!s_f+Q^2)/(2\sqrt{s_i\,s_f})$. It is related 
to an intermediate factor $\tilde{P}^0_i\,\tilde{P}^0_f $ 
that was calculated in the lab frame but whose frame dependence 
was not compensated for at later stages of the calculation. 
The present expressions fully agree with results obtained from the
simplest Feynman diagram \cite{Amghar:2002jx}, which represents 
a severe check, or with standard front-form calculations (see next
sections).}.
The normalization, $N$, is such that the charge form factor verifies 
the standard condition $F_1(0)=1$. As for the $\theta(\cdots)$ function 
and the other factors besides the wave functions $\phi(s_i), \; \phi(s_f)$,  
they result from integrating the full amplitude on the momenta 
of the on-mass-shell constituents. 
The $\theta(\cdots)$ function, in particular, arises from 
approach-dependent conditions. They include for instance the property 
that the absolute value of the cosine function of some angle should be 
smaller than 1 or the square of some momentum component should be positive.

In the simplest case of the  $F_0(Q^2)$ form factor, the expression 
to start with reads:
\begin{eqnarray}
F_0(Q^2) &\!=\!&\frac{1}{N} \int \!\!\int ds_i\;ds_f \;I(s_i,Q^2,s_f)\;
\phi(s_f) \, \phi(s_i)\, , \hspace*{2.5cm}
\end{eqnarray}
where the function $I(s_i,Q^2,s_f)$ is defined as:
\begin{eqnarray}
I(\cdots)
&\!=\!&
\frac{1}{\pi} \! \int \!\!\int \!\!\int\! 
\frac{d\vec{p}_i}{2e_i} \,\frac{d\vec{p}_f}{2e_f} \,\frac{d\vec{p}}{2e_p} \,
\delta^4(\tilde{P}_i\!-\!p_i\!-\!p)\;\delta^4(\tilde{P}_f\!-\!p_f\!-\!p) 
\,. \label{eq:def-fnI}
\end{eqnarray}
It is easy to check that the last integral, due to explicit Lorentz invariance 
of all ingredients entering the integrand, only depends on the Lorentz invariants, 
$\tilde{P}_i^2$, $\tilde{P}_f^2$, and $(\tilde{P}_i\!-\!\tilde{P}_f)^2$,
which can be identified with the Mandelstam variables $s_i$, $s_f$,  
and the squared momentum  transfer $q^2(=-Q^2)$. 
The result, whose derivation is given in appendix \ref{app:fnI}, reads:
\begin{eqnarray}
I(s_i,Q^2,s_f)&\!=\!&I
\Big(\tilde{P}_i^2,-(\tilde{P}_i\!-\!\tilde{P}_f)^2,\tilde{P}_f^2\Big)
= \frac{\theta(\cdots)}{2QD^{1/2}} \,, \label{eq:fnI}
\end{eqnarray}
where $\theta(\cdots)$ and $D$ are defined in eq. (\ref{eq:theta}).  
It is noticed that 
$\tilde{P}_i^2$ and $\tilde{P}_f^2$ are integration variables 
and differ from the squared mass of the system, $M^2$, 
which does not appear explicitly in the expression of form factors. 

It is not clear whether the quantity $D$ introduced above has some
particular meaning but it often appears in the following RQM developments.
We only remark that some could be suggested by its writing 
in terms of the above $\tilde{P}$ momenta:
\begin{eqnarray}
&&D=(\tilde{P}_i\!+\!\tilde{P}_f)^2
\!-\! \frac{\Big ((\tilde{P}_i\!+\!\tilde{P}_f)\ccdot q \Big)^2}{q^2}= 
4\frac{(\tilde{P}_i \ccdot \tilde{P}_f)^2\!-\!\tilde{P}_i^2 \tilde{P}_f^2}{Q^2}
=4\bar{s}\!+\!Q^2 \!+\! \frac{(s_i\!-\!s_f)^2}{Q^2} ,
\nonumber \\
&& {\rm with}\;\; q^{\mu}=(\tilde{P}_f\!-\!\tilde{P}_i)^{\mu} ,
 \hspace*{0.5cm} \bar{s}=\frac{s_i\!+\!s_f}{2} \,, \label{eq:sbar}
 \end{eqnarray}
where $\bar{s}$ is another notation that will also be useful later on.
While some notations appearing in expressions of the form factors, 
eqs. (\ref{eq:ff1-disp}, \ref{eq:ff0-disp}), 
refer to the original work \cite{Krutov:2002}, 
a different but equivalent writing could be useful:
\begin{eqnarray} 
F_1(Q^2) &\!=\!& \frac{1}{N} \int \!\!\int 
d\bar{s}\,d\Big(\frac{s_i\!-\!s_f}{Q}\Big) \; 
\frac{ \,\Big(2\bar{s}+Q^2\Big)\;\theta(\cdots) }{
D^{3/2} }\; \phi(s_f) \, \phi(s_i)\,, 
\label{eq:ff1-disp2}
 \\
F_0(Q^2) &\!=\!& \frac{1}{N} \int \!\! \int   
d\bar{s}\,d\Big(\frac{s_i\!-\!s_f}{Q}\Big) \; 
 \frac{\;\theta(\cdots)}{2\, D^{1/2} } \;
 \phi(s_f) \, \phi(s_i)  \, . 
\label{eq:ff0-disp2}
\end{eqnarray}
These last expressions are more condensed than the previous ones. 
Moreover, taking into account that the $\theta(\cdots)$ function 
defined by eq. (\ref{eq:theta}) puts an upper limit on  the ratio 
$\Big|\frac{s_i\!-\!s_f}{Q}\Big|$, they better evidence 
the behavior of the integrand in the limit $Q\rightarrow 0$. Accordingly, 
the two terms $s_i\!-\!s_f$ and $Q$ will be often combined to appear 
later on as in the above ratio.

\section{Constraints from covariance under space-time translations}
Lorentz invariance of form factors is a property that can be easily checked,
from boos\-ting the system for instance or, simply, from looking 
at their expressions in a few cases. Properties related to the covariance
properties of the current under Poincar\'e space-time translations, 
beyond 4-momentum conservation which is generally assumed to hold,
are not so well known, especially in the domain of RQM approaches where they
are not trivially fulfilled.

Poincar\'e covariance under space-time translations implies that a 4-vector 
current, $J^{\nu}(x)$, or a Lorentz scalar one, $S(x)$, transforms as follows:
\begin{eqnarray}
e^{iP \cdot a}\;J^{\nu}(x) \;(S(x))\;e^{-iP \cdot a}=J^{\nu}(x+a) \;(S(x+a)),
\label{transl1}
\end{eqnarray}
where $P^{\mu}$ is the operator of the Poincar\'e algebra that generates 
space-time translations. When a matrix element of the current 
between eigenstates of this operator is considered, the above relation 
allows one to factorize the space-time dependence as follows:
\begin{eqnarray}
<i\;| J^{\nu}(x) \;({\rm or}\;S(x))|\;f>=
e^{i(P_i-P_f) \cdot x}\;<i\;|J^{\nu}(0) \;({\rm or} \;S(0))|\;f>
\label{transl2}
\end{eqnarray}
Together with the factor  exp$(iq \cdot x)$ describing the external probe, 
an overall factor  exp$(i(q+P_i-P_f) \!\cdot\! x)$ is obtained. 
Invariance under space-time translations then implies the relation 
$P^{\mu}_f\!-\!P^{\mu}_i=q^{\mu} $ while the matrix element of the
current at $x=0$, at the r.h.s. of eq. (\ref{transl2}),  
can be considered as the starting point for calculating form factors.
These results have a rather general character  but they heavily 
rely on the validity of eq.~(\ref{transl1}). Fulfilling this equation 
is, most often, guaranteed in field theory whereas it could require 
some caution in the case of RQM approaches. 

With this last respect, it is interesting to consider further relations 
that can be obtained from eq. (\ref{transl1}) for small space-time displacements. 
In the simplest case, one obtains \cite{Lev:1993}:
\begin{eqnarray}
\Big[ P^{\mu}\;,\; J^{\nu}(x)\Big]=-i\partial^{\mu}\,J^{\nu}(x),
\;\;\;
\Big[ P^{\mu}\;,\; S(x)\Big]=-i\partial^{\mu}\,S(x)\, . \label{comm1}
\end{eqnarray}
Relations that could be more relevant here imply a double commutator on the
l.h.s.:
\begin{eqnarray}
\Big[P_{\mu}\;,\Big[ P^{\mu}\;,\; J^{\nu}(x)\Big]\Big]=
-\partial_{\mu}\,\partial^{\mu}\,J^{\nu}(x),
\;\;\; 
\Big[P_{\mu}\;,\Big[ P^{\mu}\;,\; S(x)\Big]\Big]=
-\partial_{\mu}\,\partial^{\mu}\,S(x)\, .\label{comm2}
\end{eqnarray}
Considering the matrix element of these last relations at $x=0$, it is found 
that the double commutator at the l.h.s. can be replaced by the square 
of the 4-momentum transferred to the system, $(P_i-P_f)^2=q^2$, 
while the derivatives at the r.h.s., for a single-particle current, 
can be replaced by the square of the 4-momentum transferred to the 
constituents, $(p_i-p_f)^2$. Poincar\'e space-time translation 
covariance, besides energy-momentum conservation already mentioned, 
therefore implies equalities like: 
\begin{eqnarray}
<\;|q^2\; J^{\nu}(0) \;({\rm or}\;S(0))|\;>=
<\;|(p_i-p_f)^2\,J^{\nu}(0)\;({\rm or}\;S(0))|\;> \,.\label{comm3}
\end{eqnarray}
In field theory, the above equations are generally fulfilled since 
the 4-momentum is conserved at the interaction vertex with the external 
probe ($p^{\mu}_f-p^{\mu}_i=q^{\mu} $). This is not necessarily true 
in RQM approaches where, under the assumption of a one-body-like component 
in the current, the squared momentum transferred to the constituents 
differs most often from that one transferred to the system: 
$(p_i-p_f)^2 \neq q^2$. Thus, a Poincar\'e-covariant calculation 
of form factors in these approaches, besides considering the matrix element 
of the current at $x=0$ at the r.h.s. of eq. (\ref{transl2}), 
also requires to  fulfill equations like eq. (\ref{comm3}), 
which involve the vicinity of $x=0$.  Fulfilling these equations 
will generally  require considering the contribution 
of many-body  components in the current.  

Looking at how  eq. (\ref{comm3}) is fulfilled in different RQM approaches,
we first notice that it is always verified in the dispersion-relation one.  
By construction, this approach  fulfills the equality 
$(p_i-p_f)^2 = q^2\;(=-Q^2)$ (see sect. 2). In the other approaches, 
which implicitly or explicitly refer to describing the physics on some 
hypersurface (hyperplane for our purpose here), eq. (\ref{comm3}) cannot be
most often satisfied, which,  for a large part,  is related to the factor
$(p_i-p_f)^2$.

Prior to calculating the quantity $(p_i-p_f)^2$, a few features 
pertinent to the general description of a two-body system 
in RQM approaches on a hyperplane have to be specified. 
The on-mass-shell momenta of the constituents, $p_1^{\mu}$ and $p_2^{\mu}$, 
are related to the total momentum of the system, $P^{\mu}$, 
the orientation of the hyperplane, $\xi^{\mu}$, 
and the off-shell invariant squared mass of the system, $s$, 
as follows \cite{Desplanques:2004sp}:
\begin{equation}
(p_1+p_2)^{\mu}=P^{\mu}
+  \Delta \; \xi^{\mu}  \, ,
\label{sump1p2} 
\end{equation}
where the quantity, $\Delta$, which represents an interaction effect, is given by:
\begin{equation}
\Delta=\frac{s-M^2}{
\sqrt{ (  P \ccdot \xi)^2 + (s\!-\!M^2)\;\xi^2 } 
+   P \ccdot \xi }\,. 
\label{Delta}
\end{equation}
The 4-vector $\xi^{\mu}$ is denoted, somewhat conventionally, 
by $\lambda^{\mu}$ when $\xi^2$ is finite (most often, one then takes 
$\lambda^2=1$ though the results should be independent of this value) 
or by $\omega^{\mu}$ when $\xi^2$ is equal to zero (front-form case). 
The above definitions, together with $P^2=M^2$, allow one to verify 
the relation:
\begin{equation}
s=(p_1+p_2)^2\, ,
\label{4ek2} 
\end{equation}
indicating that the variable $s$ can be identified to that one introduced 
in the dispersion-relation approach previously described. On the other hand, 
$s$ is most often expressed in terms of the internal momentum variable, 
$\vec{k}$, which enters the mass-operator description ($\vec{k}$ can be 
considered as a relativistic generalization of the c.m.  momentum carried 
by the constituents in the instant form \cite{Bakamjian:1953kh}). 
The relation, $s=4\,e^2_{k}=4\,(m^2+k^2)$,
thus suggests that the wave function, $\phi(s)$, which appears 
in the expression of form factors in the dispersion-relation approach, 
eqs. (\ref{eq:ff1-disp}, \ref{eq:ff0-disp}), and the solution of the mass operator, 
$\tilde{\phi}(k^2)$ 
(with normalization $8\int dk\,k^2/e_k \,\tilde{\phi}^2(k^2)=N$), 
could be related as follows:
\begin{equation}
\phi(s)=\tilde{\phi}(\frac{s-4m^2}{4})=\tilde{\phi}(k^2)\, .
\end{equation}
For practical purposes, we assign to the constituent 1 the role 
of the constituent interacting with the external field. 
Its momentum is denoted $p_i^{\mu}$ or $p_f^{\mu}$ depending 
it refers to the initial or final states. The constituent 2 is
assigned the role of the spectator and its momentum is denoted $p^{\mu}$.  

Taking into account that one of the constituent is a spectator, 
the difference $(p_i-p_f)^{\mu}$ can be expressed as follows:
\begin{equation}
(p_i\!-\!p_f)^{\mu} =(P_i\!-\!P_f)^{\mu}+ (\Delta_i \!-\!\Delta_f)\,\xi^{\mu}\, .
\label{difpipf}
\end{equation}
We assume that the initial and final states are described on a unique 
hyperplane (in the ``point form" of Bakamjian \cite{Bakamjian:1961} 
and Sokolov \cite{Sokolov:1985jv}, 
which involves hyperplanes perpendicular to the velocity of the system, 
two orientations should be introduced, which is manageable). 
From eq. (\ref{difpipf}), one now gets:
\begin{eqnarray}
(p_i\!-\!p_f)^2&=&
(P_i\!-\!P_f)^2+2\, (\Delta_i \!-\! \Delta_f)\;  (P_i\!-\!P_f) \ccdot \xi
+(\Delta_i \!-\!\Delta_f)^2 \;\xi^2
\nonumber \\
&=&q^2-2\,(\Delta_i \!-\!\Delta_f) \;q \ccdot \xi +(\Delta_i \!-\!\Delta_f)^2\;\xi^2 \,.
\label{difpipf2}
\end{eqnarray}
Examination of this last equation shows that the interaction 
effects represented by the $\Delta$ quantity generally prevent one 
from fulfilling the equality given by eq. (\ref{comm3}). For a pion-like system,
this one can be violated by more or less large factors, depending 
on the approach and how much the system is bound. 
It is thus found in ref. \cite{Desplanques:2004sp}  
that the largest violation was observed for the ``point-form" approach
where it reaches a factor 35000 at $Q^2=100 \,({\rm GeV/c})^2$. The smallest
violation was observed  for the standard instant form (Breit frame) 
where it amounts to a few \% and for the standard front form $(q^+=0)$ 
where it is absent. In this case, the result can easily be seen from 
eq. (\ref{difpipf2}). The first order term in $\Delta$ does not contribute 
due to the condition $\xi \!\cdot\!q=q^+=0$ (this also holds for the standard 
instant form). The second-order term does not contribute either due to the
condition $\xi^2=0$ (also verified in the instant form with $\xi^2=1$ 
together with $|\vec{P}_i+\vec{P}_f| \rightarrow \infty$ and $E_i=E_f$). 
Thus, the standard front-form approach is expected 
to play a particular role in the comparison with the dispersion-relation one.

\section{Implementing constraints in RQM approaches}
Examination of expressions for  form factors shows that the discrepancy 
between different approaches can be largely ascribed to the factor 
that multiplies the momentum transfer, $q$, on which they depend. 
In the most striking cases, this factor takes the form $2e_k/M$, 
where $2e_k$ represents the internal kinetic energy of the system 
(see sect. 4.1.2 in ref. \cite{Desplanques:2004} for a related example). 
As it can be infered from 
$\Delta \propto s-M^2 (=4e_k^2-M^2)$,
a similar factor is
responsible to make the squared momentum transfer to the constituents, 
$(p_i-p_f)^2$ different from that one transferred to the system, $q^2$. 
Typically, this factor deviates from 1 by terms that have an interaction
character and are here or there depending on the approach. 
This observation suggests that multiplying $q$ by a factor $\alpha$, 
so that the above relation be fulfilled, could account for missing interaction
effects in the approach under consideration. This amounts to account for the
contribution of many-body currents which are expected to ensure 
the equivalence of different approaches. In the next subsections, 
we proceed with the derivation of this factor in the instant form, 
the front form and the ``point form", 
with arbitrary momentum configurations in all cases. 
In this order, we are led to introduce the average momentum carried 
by the system, $ \bar{P}^{\mu}=\frac{1}{2}( P_i^{\mu}\! +\! P_f^{\mu})$, 
which implies the relations 
$P_i^{\mu}=\bar{P}^{\mu}\!-\!\frac{1}{2}q^{\mu}$, 
$P_f^{\mu}=\bar{P}^{\mu}\!+\!\frac{1}{2}q^{\mu}$.
While doing so, we faced the question of whether  $ \bar{P}^{\mu}$ 
should be considered as an independent variable or a variable depending 
on $q$ (notice that $\bar{P}^2=M^2\!-\!q^2/4$). The choice retained here has been motivated 
by finding a tractable solution for the factor $\alpha$. We don't therefore
exclude other solutions. At least, one was found in a very particular case 
(front form in the Breit frame with the momentum transfer perpendicular 
to the front orientation). The corresponding expression for $\alpha$ 
and the related expression of the charge form factor are given 
in appendix \ref{gen:other}. As can be seen there, they are not characterized 
by their simplicity but we believe that the existence of other solutions 
is conceptually important. This freedom could be used to fulfill 
further relations due to considering commutators different 
from the double one, eq. (\ref{comm2}), which  is sufficient 
for the present work. 

\subsection{Determination of $\alpha$ in the instant form}
In order to determine the factor $\alpha$ in the instant form, 
we directly start from the expression of $(p_i\!-\!p_f)^2$ where 
the momenta $p_i^{\mu}$ and $p_f^{\mu}$ are expressed in terms of the spectator 
momentum $p^{\mu}$ and the total momenta $P_i^{\mu}$ and $P_f^{\mu}$, 
taking into account eq. (\ref{sump1p2}) and the definition of $\lambda^{\mu}$ 
in the instant form, $\lambda^{\mu}=(1,\, 0,\,0,\,0)$: 
\begin{eqnarray}
(p_i\!-\!p_f)^2&=&(e_i \!-\!e_f)^2 -(\vec{p}_i\!-\!\vec{p}_f)^2
=\Bigg (\sqrt{m^2\!+\!(\vec{P}_i\!-\!\vec{p}\,)^2}
- \sqrt{m^2\!+\!(\vec{P}_f\!-\!\vec{p}\,)^2} \Bigg )^2 -\vec{q}\,^2 
\nonumber \\
&=&\Bigg (\sqrt{m^2+(\vec{\bar{P}}-\frac{1}{2}\vec{q}-\vec{p}\,)^2}
- \sqrt{m^2+(\vec{\bar{P}}+\frac{1}{2}\vec{q}-\vec{p}\,)^2}\Bigg)^2 -\vec{q}\,^2 \,.
\end{eqnarray}
As expected, the above quantity differs from the squared transfer momentum 
$q^2=q_0^2-\vec{q}\,^2=-Q^2$. Following the procedure described 
in the introduction of this section, we multiply  $\vec{q}$ 
in the above expression by the factor $\alpha$ and require 
the equality $(p_i-p_f)^2=q^2$ be fulfilled. Actually, the change also
incorporates a change from $|\vec{q}|$ to $\sqrt{\vec{q}\,^2-q_0^2}$, 
which can be absorbed into the definition of the factor $\alpha$. 
We thus get the equation:
\begin{eqnarray}
\Bigg (\sqrt{m^2\!+\!
(\vec{\bar{P}}\!-\!\hat{q}\,\frac{\alpha\sqrt{\!-q^2}}{2}\!-\!\vec{p}\,)^2}
- \sqrt{m^2\!+\!
(\vec{\bar{P}}\!+\!\hat{q}\,\frac{\alpha\sqrt{\!-q^2}}{2}\!-\!\vec{p}\,)^2}\Bigg)^2
 +\alpha^2\,q^2 =q^2\,,
\end{eqnarray}
where  $\hat{q}$ is a standard unit 3-vector defined as 
$\hat{q}= \vec{q}/|\vec{q}| $. From the above equation, we obtain the solution:
\begin{eqnarray}
\alpha^2=\frac{m^2\!+\!(\vec{\bar{P}}\!-\!\vec{p}\,)^2\!-\!\frac{q^2}{4} }{ 
m^2\!+\!(\vec{\bar{P}}\!-\!\vec{p}\,)^2\!-\!
((\vec{\bar{P}}\!-\!\vec{p}\,)\ccdot \hat{q})^2\!-\!\frac{q^2}{4}}
=1+\frac{((\vec{\bar{P}}\!-\!\vec{p}\,) \ccdot \hat{q})^2}{m^2\!+\!
(\vec{\bar{P}}\!-\!\vec{p}\,)^2 \!-\!((\vec{\bar{P}}\!-\!\vec{p}\,) \ccdot \hat{q})^2
\!-\!\frac{q^2}{4}} \,.
\label{alpif}
\end{eqnarray}
Simpler expressions can be obtained for a parallel or a perpendicular
kinematics, $\vec{q} \parallel \vec{\bar{P}}$ and  
$\vec{q} \perp \vec{\bar{P}}$  respectively. They can be helpful 
to simplify the foregoing developments, especially for the parallel 
kinematics. The Breit-frame case, which is common to these two
kinematics, is considered explicitly here. 

A generalization of the above results to an arbitrary orientation 
of the hyperplane defined by the 4-vector $\lambda^{\mu}$ may be considered.
As it can be inferred from the above instant-form approach by applying 
a boost transformation determined by the 4-vector velocity 
$v^{\mu}=\lambda^{\mu}$, this generalization does not present 
major difficulty. Some detail can be found in appendix \ref{gen:hyp}.

\subsection{Determination of $\alpha$ in the front form}
The front-form approach is characterized by an hyperplane whose orientation, 
$\omega^{\mu}$, fulfills the condition $\omega^2=0$. Using the relation 
between the momenta of the constituents, the total momentum 
and the above orientation, eq. (\ref{sump1p2}), we can write 
the quantity $(p_i\!-\!p_f)^2$ as:
\begin{eqnarray}
(p_i\!-\!p_f)^2\!&\!=\!&\!q^2\!-q\!\cdot\!\omega
\Bigg (\frac{m^2\!-\!(\bar{P}\!-\!\frac{q}{2}\!-\!p)^2  }{ 
(\bar{P}\!-\!\frac{q}{2}\!-\!p) \!\cdot\! \omega}
-\frac{m^2\!-\!(\bar{P}\!+\!\frac{q}{2}\!-\!p)^2  }{ 
(\bar{P}\!+\!\frac{q}{2}\!-\!p)\!\cdot\! \omega}\Bigg)\,.
\end{eqnarray}
Assuming that all the components of $q^{\mu}$ are affected in
the same way, the equation to be fulfilled by $\alpha$ is given by:
\begin{eqnarray}
\alpha^2q^2\!-\alpha^2q\!\cdot\!\omega \Bigg (
\frac{(-\bar{P}^2\!+\!2p\!\cdot\!\bar{P}\!-\!\alpha^2\frac{q^2}{4})(q\!\cdot\!\omega)
+2(\bar{P}\!-\!p)\!\cdot\! q \;(\bar{P}\!-\!p)\!\cdot\! \omega)
}{(\bar{P}\!-\!p)\!\cdot\! \omega)^2- \alpha^2(\frac{q}{2}\!\cdot\! \omega)^2}
\Bigg)=q^2\,.
\end{eqnarray}
Its solution reads: 
\begin{eqnarray}
\alpha^2=\frac{1}{1 
-2 \frac{q \cdot \omega}{(\bar{P}\!-\!p) \cdot \omega}
\frac{(\bar{P}\!-\!p) \cdot q }{q^2}+
\frac{(q \cdot \omega)^2}{((\bar{P}\!-\!p) \cdot \omega)^2}
\frac{\bar{P}^2\!-\!2p \cdot \bar{P}\!+\!\frac{q^2}{4}}{q^2}}\,.
\label{alpff}
\end{eqnarray}
It is noticed that the factor $\alpha$ reduces to 1 when the condition 
$q \!\cdot\! \omega =0$ (often referred to as $q^+=0$) is fulfilled.

\subsection{Determination of $\alpha$  in the ``point form"}
The ``point form" under consideration here is an instant form with the symmetry
properties of the Dirac point form \cite{Bakamjian:1961}. As mentioned by
Sokolov, it supposes physics described on a hyperplane perpendicular to the
velocity of the system \cite{Sokolov:1985jv}. In calculating form factors, 
two hyperplanes are therefore implied. Thus restoring constraints stemming 
from space-time translation invariance in this approach may evidence features
different from the other ones previously considered. Moreover, as noticed by
Coester \cite{Coester:2003zh}, the constituent momenta, $p, \; p_i, \;p_f$, 
do not generate in the present case translations consistent with the dynamics. 
A point-form approach \cite{Desplanques:2004rd}, more in the spirit 
of the Dirac one, based on a hyperboloid surface,  is mentioned 
at the end of sect. \ref{ssec:ff}.

For our purpose, we start from the square of the momentum transfer 
to the struck constituent. Contrary to the other forms, the consideration 
of the inelastic case requires however some caution. For simplicity, 
we therefore only consider here the elastic case ($M_i^2=M_f^2=M^2$). The
corresponding expression of $(p_i\!-\!p_f)^2$ can thus be written as:
\begin{eqnarray}
(p_i\!-\!p_f)^2=(e_i\!-\!e_f)^2- (\vec{p}_i\!-\!\vec{p}_f)^2 
=\frac{4}{M^2} \Big(1\!-\!\frac{q^2}{4M^2}\Big)
\Bigg((p \!\cdot\! q)^2 +q^2(p \!\cdot\! \hat{v})^2\Bigg) \, ,
\end{eqnarray}
where $p^{\mu}$ represents the spectator on-mass-shell 4-momentum and 
$q^{\mu}$ the momentum transfer $ (P_f\!-\!P_i)^{\mu}$. The unit 4-vector 
$\hat{v}^{\mu}$ represents the overall velocity of the system defined as 
$ (P_f\!+\!P_i)^{\mu} /\sqrt{(P_f\!+\!P_i)^2}$. Noting that 
$(P_f\!+\!P_i)^2=4M^2+Q^2$, its components may depend on the square momentum 
transfer $q^2=-Q^2$. In the following however, this 4-vector  
is treated as if it was not depending on this variable, quite in the spirit 
of the point-form  approach where velocities can be considered 
as independent variables (see construction of the Poincar\'e algebra). 
This detail, which is form dependent, is important in determining 
the equation that has to be solved to get the factor $\alpha$. 
In accordance with the expected Lorentz invariance of form factors 
in the approach under consideration here, we assume that the factor 
$\alpha$ affects in the same way all the components of $q^{\mu}$. The
equation is thus given by:
\begin{eqnarray}
\frac{4\alpha^2}{M^2} \Big(1-\alpha^2\frac{q^2}{4M^2}\Big)
\Bigg((p \!\cdot\! q)^2 +q^2(p \!\cdot\!\hat{v} )^2\Bigg)=q^2 \,.
\end{eqnarray}
Its solution is given by:
\begin{eqnarray}
\alpha^2=\frac{M^2}{2\,\sqrt{(p \!\cdot\! \hat{v})^2 \!-\! (p \!\cdot\! \hat{q})^2  } \;\Big(
\sqrt{(p \!\cdot\! \hat{v})^2 \!-\! (p \!\cdot\! \hat{q})^2}
+\sqrt{(p \!\cdot\! \hat{v})^2 \!-\! (p \!\cdot\! \hat{q})^2\!+\!\frac{Q^2}{4}}\Big)}
\,,
\label{alppf}
\end{eqnarray}
where $\hat{q}^{\mu}=q^{\mu}/\sqrt{-q^2}$. Notice that the definition of
$\hat{q}^{\mu}$ involves here a 4-vector 
(instead of a 3-vector in the instant form) with the consequence 
that $\hat{q}^2=-1$ (instead of $\hat{q}^2=1$  in the instant form). 
Due to the Lorentz
invariance property of the approach, the above results are a rather
straightforward generalization of those obtained from the Breit frame, with a
boost transformation represented by the 4-vector $\hat{v}^{\mu}$.

%
\section{Expressions of form factors in the Breit frame with constraints 
implemented} \label{sec:rqm}
We here show how the constraints discussed in the previous section affect 
the calculation of form factors and, in particular, allow one to recover 
the dispersion-relation expressions, provided that the current 
is appropriately chosen. This is done for various forms. 
We outline below the main steps for the Breit-frame case. 
For convenience, and in absence of ambiguity, we now refer most often
to the notation $Q$ instead of $q$ for the momentum transfer ($Q^2$ is always
positive for elastic form factors considered here).
Some detail  as well as generalizations to arbitrary frames 
could be found in appendices \ref{app:detail} and \ref{app:gen} respectively. 

In all cases, we begin by expressing the quantities $s_i, s_f$ 
in terms of the spectator momentum, $\vec{p}$, together with the momenta 
of the initial and final states. We then invert these relations so that to
express the spectator momentum (and related quantities) in terms 
of  $s_i, s_f$ that enter dispersion-relation expressions and a third variable 
that has to be integrated over independently. 
Schematically, the expected result takes the form:
\begin{eqnarray}
\int \frac{d\vec{p}}{e_p} \;
 ^{``}\Big(
\tilde{\phi}(\vec{k_f}^2)\;\tilde{\phi}(\vec{k_i}^2)\cdots\Big)^{"} 
 = \int\! \!\int ds_i \, ds_f\, \phi(s_f)\;\phi(s_i)\cdots
 \int dz \,f(z,\cdots)\, ,
\end{eqnarray} 
where the notation ``$\;$" indicates that the inserted quantity 
accounts for the effect of constraints related to space-time translation 
invariance, as described in appendix \ref{app:detail}, 
while the dots stand for the current as well as the approach under
consideration. The choice of the last variable, $z$,  
is not unique and some could be more appropriate, depending for instance 
on the approach or on the symmetry properties of the momentum configuration. 
In any case, this is an important and non-trivial step. In the general case, 
this is the place where possible dependence of intermediate
ingredients on non-Lorentz invariant quantities can be absorbed 
in the variable so that to  disappear after it is integrated over.

\subsection{Instant form} \label{ssec:if}
The quantities $s_i, s_f$ we start from are given in appendix
\ref{detail:if}. Due to the azimuthal symmetry of the momentum configuration
considered here, it is appropriate to express them in terms of 
$p^2_{\perp}$ and $p_{\parallel}$ where $p_{\perp}$ and $p_{\parallel}$ 
represent the components of $\vec{p}$  perpendicular and parallel
to the momentum transfer $\vec{q}$. For convenience, 
we use the combinations $\bar{s}=\frac{s_i\!+\!s_f}{2}$ and $(s_i\!-\!s_f)$,
which are respectively symmetrical and  antisymmetric 
in the exchange of initial and final states\footnote{This choice 
offers the advantage to partly disentangle the component of the spectator momentum 
along the momentum transfer. The factor $\frac{1}{2}$ has been introduced 
so that the change of these combinations back to the original variables 
does not introduce extra factors in the integration volumes 
($d\bar{s} \;d(s_i\!-\!s_f)= ds_i\,ds_f$).}: 
\begin{eqnarray}
&&\bar{s}=\frac{s_i\!+\!s_f}{2}= 2\,\sqrt{m^2\!+\!p^2_{\perp}\!+\!p^2_{\parallel}}\,
\Bigg(\sqrt{m^2\!+\!p^2_{\perp}\!+\!p^2_{\parallel}}
+\sqrt{m^2\!+\!p^2_{\perp}\!+\!p^2_{\parallel}\!+\!\frac{1}{4}\,Q^2}  \Bigg)\,,
\nonumber \\
&&s_i\!-\!s_f=2\,\frac{p_{\parallel}\;Q}{\sqrt{m^2\!+\!p^2_{\perp}\!+\!\frac{1}{4}\,Q^2}}
\Bigg(\sqrt{m^2\!+\!p^2_{\perp}\!+\!p^2_{\parallel}}
+\sqrt{m^2\!+\!p^2_{\perp}\!+\!p^2_{\parallel}\!+\!\frac{1}{4}\,Q^2}  \Bigg)\, .
\end{eqnarray} 
Inverting these relations, we obtain:
\begin{eqnarray}
 p^2_{\perp}=
 \frac{s_i\;s_f}{D} -m^2\, , \hspace*{1cm}
p_{\parallel} =
 \frac{s_i\!-\!s_f}{Q}\; \frac{2\bar{s} +Q^2   }{  
 2\,\sqrt{D}  \;  \sqrt{4\bar{s}\!+\!Q^2} }\, ,
\label{chgif}
\end{eqnarray}
where the positivity of $p^2_{\perp}$ allows one to recover the condition
implied by the $\theta(\cdots)$ function given previously, 
eq. (\ref{eq:theta}).

Noticing that $d\vec{p}=dp_{\perp}^2\, dp_{\parallel}\,\frac{d\phi}{2} $, 
it is found that the transformation of the $\vec{p}$ variable 
to the variables $s_i, s_f,$ and the extra one ($\phi$ in the present case), 
implies the  following relation between the integration volumes:
\begin{equation}
d\vec{p}\; \frac{``(e_i\!+\!e_f\!+\!2\,e_p)"}{2e_p\,``(e_i\!+\!e_f)"}=
 d\bar{s} \; d( \frac{s_i\!-\!s_f}{Q})\;  
\frac{ (2\bar{s}+Q^2)  }{ 2\,D^{3/2}}\;
\frac{d\phi}{2}\,.
\end{equation}

The factor $\frac{``(e_i\!+\!e_f\!+\!2\,e_p)"}{2\,``(e_i\!+\!e_f)"}$ 
can be considered as part of the charge current and ensures that the
charge is independent of the velocity of the system. It  differs 
from that one naively expected for an interaction of the external probe 
with positive-energy constituents, as generally assumed in RQM approaches 
\cite{Amghar:2002jx}. 
In the present case, this assumption which amounts to neglect contributions
corresponding to Z-type diagrams, implies a current with time component 
proportional to $e_i\!+\!e_f$. Actually, the above factor was obtained 
from an analysis involving the triangle Feynman
diagram \cite{Amghar:2002jx} but, as it had a small effect, it was
discarded in numerical applications. It is however relevant if one aims 
to establish the equivalence of different approaches.
Putting together the above expression and the other ingredients entering
the charge form factor,  one gets: 
\begin{eqnarray} 
``F_1(Q^2)"&\!=\!&\frac{16\pi^2}{N}\! \int\! \frac{d\vec{p}}{(2\pi)^3}\;
\frac{1}{e_p} \;^{``}\Bigg( \frac{(e_f+e_i+2e_p)}{2\,(e_f+e_i)} \;
\tilde{\phi}(\vec{k_f}^2)\;\tilde{\phi}(\vec{k_i}^2) \Bigg)^{"}
\nonumber \\
&\!=\!&\frac{1}{\pi\,N}\! \int \!\! \int  d\bar{s} \; d( \frac{s_i\!-\!s_f}{Q})\; \; 
 \frac{ (2\bar{s}+Q^2) \; \theta(\cdots)  }{ D^{3/2}} \;
\phi(s_f)\;\phi(s_i) \times \! \int \!\frac{d\phi}{2}\, .
\label{ifx1} 
\end{eqnarray} 
Taking into account that the last integral is equal to $\pi$, 
the expression of $F_1(Q^2)$ given by eq. (\ref{eq:ff1-disp2}) 
is easily recovered.

While the expression for the charge form factor we started from is not
unexpected, there is no similar result for the scalar form factor.
In this case, some trick motivated by the desired result can be used. 
It consists in multiplying the integrand for the charge form factor 
by the factor $\frac{D}{2(2\bar{s}+Q^2)}$ where the quantities
$s_i$, $s_f$ and $q^2(=-Q^2)$ are respectively replaced 
by $(p_i\!+\!p)^2$, $(p_f\!+\!p)^2$ and $(p_i\!-\!p_f)^2$. 
This procedure gives the following result:
\begin{eqnarray} 
``F_0(Q^2)"\!&\!=\!&\!\frac{16\pi^2}{N}\! \int\! \frac{d\vec{p}}{(2\pi)^3}\;
\frac{1}{e_p} 
\; ^{``}\Bigg(\Bigg(\frac{e_i\!+\!e_f\!+\!2e_p}{2\,(e_i\!+\!e_f)}\Bigg)^2\,
 \Bigg(\!1\!+\!\frac{(e_i\!-\!e_f)^2}{(e_i\!+\!e_f)^2\!-\!4e_p^2}\Bigg)\;
\tilde{\phi}(\vec{k_f}^2)\;\tilde{\phi}(\vec{k_i}^2)\Bigg)^{"}
\nonumber \\ 
\!&\!=\!&\!\frac{1}{\pi\,N}\! \int \!\!\int d\bar{s} \; d( \frac{s_i\!-\!s_f}{Q})\; 
 \frac{\;\theta(\cdots) }{2\,D^{1/2}} 
\;\phi(s_f)\;\phi(s_i) \times \! \int \frac{d\phi}{2}\, .
\label{ifx0} 
\end{eqnarray} 
Examining the first line of the above expression, 
especially the unexpected factor  $e_i\!+\!e_f\!-\!2e_p$
at the denominator,  it is not obvious 
whether one can find some field-theory justification. 
This is perhaps the indication that the trick we used, 
which can be employed without any modification in other cases,
is a too rough one. With this respect, we notice that the comparison 
with the triangle Feynman-diagram contribution suggests a dependence 
of the integrand on the total mass of the system, which is absent here.

\subsection{Front form} \label{ssec:ff}
Contrary to the instant form or the ``point form" (see below) 
with the Breit frame, the front form with the same momentum configuration 
involves two directions, the orientation of the momentum transfer, 
$\hat{q}=\vec{q}/|\vec{q}|$ and of the front,  
$\hat{n}=\vec{\omega}/\;\omega^0$. Lacking the azimuthal symmetry 
present in the other cases, the corresponding expressions of  $s_i, s_f$, 
given in appendix \ref{detail:ff}  are generally more complicated. 
As a consequence, form factors cannot be simply reduced to a two-dimensional 
integral.  In terms of the quantities $\vec{p} \cdot \hat{n} $, 
$\vec{p} \cdot \hat{q} $,  $\hat{q} \cdot \hat{n} $, $Q=|\vec{q}|$ 
and $\bar{E}^2=M^2+Q^2/4$, the expressions of interest here, $\bar{s}$ 
and $s_i\!-\!s_f$, read:
\begin{eqnarray}
&&\bar{s}=
2\frac{\bar{E}}{\bar{E}\!-\!e_p\!+\!\vec{p} \ccdot \hat{n}} 
\; e_p\bar{E}
-\frac { e_p \!-\! \vec{p} \ccdot \hat{n}  }{ 
\bar{E} \!-\! e_p \!+\! \vec{p} \ccdot \hat{n}  }\; M^2\,,
\nonumber \\
&&s_i\!-\!s_f=2Q\;
\frac { \frac{\bar{E}}{\bar{E}\!-\!e_p\!+\!\vec{p} \cdot \hat{n}}\; 
\vec{p} \ccdot \hat{q}
-\frac { e_p \!-\! \vec{p} \cdot \hat{n}  }{ 
2(\bar{E} \!-\! e_p \!+\! \vec{p} \cdot \hat{n})  }
\frac{\hat{q} \cdot \hat{n}}{\bar{E} \!-\! e_p \!+\! \vec{p} \cdot \hat{n}}\;
 (2e_p \bar{E} \!-\! M^2  )
}{ \sqrt {1 
-2 \frac{\hat{q} \cdot \hat{n}}{\bar{E} \!-\! e_p \!+\! \vec{p} \cdot \hat{n}}
\;\vec{p} \ccdot \hat{q}
+\frac{(\hat{q} \cdot \hat{n})^2  }{  
(\bar{E} \!-\! e_p \!+\! \vec{p} \cdot \hat{n})^2}\;(2e_p \bar{E} \!-\! M^2  )
 } }\,. \label{eq:rqm-ff1}
\end{eqnarray}
Assuming that the vectors $\hat{n}$ and $\hat{q}$ define a $x,\,y$ plane, 
we make the change of the variables $p^x,\,p^y,\,p^z$ 
into the variables $\bar{s}$, $s_i \!- \! s_f$ and $p^z$ ultimately.
The above expressions are inverted first to express the quantities 
$\vec{p} \!\cdot\! \hat{n} $ and $\vec{p} \!\cdot\! \hat{q} $ in terms of 
$\bar{s}$, $s_i \!- \! s_f$  and $e_p$:
\begin{eqnarray}
&&\vec{p} \ccdot \hat{n}=\frac{e_p(2\bar{E}^2\!+\!\bar{s}\!-\!M^2)-\bar{s}\bar{E} 
}{\bar{s}\!-\!M^2}\, ,
\nonumber \\
&&\vec{p} \ccdot \hat{q}=
\frac{\hat{q} \!\cdot\! \hat{n}}{\bar{E}}\;\frac{\bar{s}\!-\!2e_p\bar{E}}{2}
+\frac{2e_p\bar{E}\!-\!M^2}{2(\bar{s}\!-\!M^2)}(\frac{s_i\!-\!s_f}{Q})
\Big(\sqrt{E2}\!-\!\frac{\hat{q} \!\cdot\! \hat{n}}{\bar{E}}
\frac{s_i\!-\!s_f}{2Q}\Big)\, , \label{eq:rqm-ff2}
\end{eqnarray}
where 
$E2=1 \!+\!\Big (\frac {\hat{q} \cdot \hat{n}}{ \bar{E} }\Big)^2
\Big (\bar{s}\!-\!M^2\!+\! \frac {(s_i\!-\!s_f)^2}{4Q^2}\Big)$.
Using these last expressions to determine the components $p^x$ and $p^y$, 
together with the relation $e_p^2=m^2\!+\!p^{x2}\!+\!p^{y2}\!+\!p^{z2}$,  
one then determines the expression of $e_p$ in terms of   $s_i, s_f$ 
and $p^z$: 
\begin{eqnarray}
&&e_p=\frac{M^2}{2\bar{E}} +\frac{\bar{s}\!-\!M^2}{2\bar{E}} \,
\frac {\sqrt{\frac{E2}{D}}\,(2\bar{s} \!+\!Q^2 ) \pm 2
\sqrt{1\!-\!(\hat{q} \ccdot \hat{n})^2}
\sqrt{\frac{s_is_f}{D}\!-\!(m^2\!+\!p^{z2})} 
}{ \sqrt{D}\Big(\sqrt{E2}\!-\!\frac{\hat{q} \cdot \hat{n}}{\bar{E}}\,
\frac{s_i\!-\!s_f}{2Q}\Big)}\,. \label{eq:rqm-ff3}
 \end{eqnarray}
The integration volume can now be expressed in terms of the new variables. It
fulfills the relation:  
\begin{eqnarray}
&& \hspace*{-0.5cm}\frac{d\vec{p}}{e_p} \; 
\frac {\bar{E}}{2 (\bar{E}\!-\!e_p\!+\!\vec{p} \ccdot \hat{n}) } \;
=\sum \frac{1}{4}\frac {d\bar{s} \; d(\frac{s_i\!-\!s_f}{Q})\, dp^z  }{  
 \sqrt{D} \sqrt{ \frac{s_is_f}{D}\!-\!(m^2\!+\!p^{z2})} } 
\nonumber \\ 
&& \hspace*{0.5cm} \times 
\Bigg ( \frac{(2\bar{s}\!+\!Q^2)
 \pm 2\sqrt{\frac{D}{E2}}\sqrt{1\!-\!(\hat{q} \ccdot \hat{n})^2}
   \sqrt{\frac{s_is_f}{D}\!-\!(m^2\!+\!p^{z2})} }{ D} \Bigg)  \,,
    \label{eq:rqm-ff4}
\end{eqnarray}
where the sum symbol reminds that the two solutions for $e_p$ 
in eq. (\ref{eq:rqm-ff3}) should be accounted for. 
Similarly to the instant form case, an expression 
that fulfills minimal pro\-per\-ties is obtained for the charge form factor: 
\begin{eqnarray} 
``F_1(Q^2)"&\!=\!&\frac{16\pi^2}{N}\! \int\! \frac{d\vec{p}}{(2\pi)^3}\;
\frac{1}{e_p} \;
^{``}\Bigg(\frac {\bar{E}}{2 (\bar{E}\!-\!e_p\!+\!\vec{p} \ccdot \hat{n}) } \;
\tilde{\phi}(\vec{k_f}^2)\;\tilde{\phi}(\vec{k_i}^2) \Bigg)^{"}
\nonumber \\
&\!=\!&\frac{2}{\pi N}\! \int\!\! \int d\bar{s} \; d(\frac{s_i\!-\!s_f}{Q})\;
  \frac{ \theta(\cdots)  }{
4D^{3/2}} 
\;\phi(s_f)\;\phi(s_i)
\nonumber \\
&& \hspace*{1.0cm} \times\!\sum
\int\! dp^z   
  \frac{(2\bar{s}\!+\!Q^2)\pm 2\sqrt{\frac{D}{E2}}
\sqrt{1\!-\!(\hat{q} \ccdot \hat{n})^2}
\sqrt{\frac{s_is_f}{D}\!-\!(m^2\!+\!p^{z2})} }{ 
 \sqrt{ \frac{s_is_f}{D}\!-\!(m^2\!+\!p^{z2})} }\, .\label{eq:rqm-ff5}
\end{eqnarray} 
While performing the sum over the two values of $e_p$ at the last line,
it is noticed that the total result becomes independent of the front 
orientation,  $\hat{n}$, which appears in the factor $\frac{1}{\sqrt{E2}}$, 
thus contributing to its rotation invariance. Moreover, it becomes independent
of the mass of the system, $M$. Making the integration over $p^z$ 
then provides a factor $2\pi(2\bar{s}\!+\!Q^2)$, allowing one to recover eq.
(\ref{eq:ff1-disp2}).

Results of this section could apply to the Dirac point form, 
as far as the calculation of form factors in this approach 
\cite{Desplanques:2004rd} amounts to consider an appropriately weighted 
superposition of contributions corresponding to hyperplanes with continuously 
varying orientations and $\omega^2=0$ \cite{Desplanques:2004sp}.

\subsection{``Point form"} \label{ssec:pf}
The appropriate expressions of   $s_i, s_f$ to be used in the ``point-form" 
case are derived in appendix \ref{detail:pf}. Due to the azimuthal symmetry 
of the problem, 
they are most easily expressed in terms of the variables $p^2_{\perp}$ 
and $p_{\parallel}$, already defined in the instant-form case. 
The combinations with a symmetry character are given by:
\begin{eqnarray}
&&\bar{s} \!=\! 2\Bigg[ \sqrt{m^2\!+\!p^2_{\perp}}\,
\Bigg(\!\sqrt{m^2\!+\!p^2_{\perp}}\!+\! \sqrt{m^2\!+\!p^2_{\perp}\!+\!\frac{Q^2}{4}}
   \Bigg) \!+\!\frac{2p^2_{\parallel}}{\sqrt{m^2\!+\!p^2_{\perp}}} 
   \sqrt{m^2\!+\!p^2_{\perp}\!+\!\frac{Q^2}{4}} \Bigg]\, ,
\nonumber \\
&&s_i\!-\!s_f  \!=\!
4\;\frac{p_{\parallel}\;Q\;\sqrt{m^2\!+\!p^2_{\perp}\!+\!p^2_{\parallel}}}{
\sqrt{m^2\!+\!p^2_{\perp}}}\,.
\label{eq:ek2pf}
\end{eqnarray} 
Inverting these relations, one gets:
\begin{eqnarray}
 &&p^2_{\perp}=
 \frac{s_i\;s_f}{D} -m^2\, , \hspace*{1cm}
p_{\parallel} =
 (\sqrt{s_i}\!-\!\sqrt{s_f})\; \frac{ (s_i\;s_f)^{1/4} }{2\;
 \sqrt{(\sqrt{s_i}\!-\!\sqrt{s_f})^2+Q^2}}\, ,
\end{eqnarray}
where, again, the positivity of $p^2_{\perp}$ allows one to recover 
the condition implied by the $\theta(\cdots)$ function given previously, 
eq. (\ref{eq:theta}). The above transformation implies 
the  following relation between the integration volumes:
\begin{eqnarray}
\frac{d\vec{p}}{e_p}=\frac{dp^2_{\perp}}{e_p}\; dp_{\parallel}\,
\frac{d\phi}{2}&=&
\frac{d\bar{s}\; d(s_i\!-\!s_f)\;(2\bar{s}\!+\!Q^2)}{2\,Q\,D^{3/2}}\,
\frac{d\phi}{2}\, .\label{eq:jac-pf} 
\end{eqnarray}
Together with the other ingredients, this expression allows one to write 
the charge form factor as:
\begin{eqnarray} 
``F_1(Q^2)" & =  & \frac{16\,\pi^2}{N} \; \int \frac{d \vec{p}}{(2\pi)^3} \; 
\frac{1}{e_p} \;
^{``}\Bigg(\tilde{\phi}(\vec{k}_{f}^2) \;\tilde{\phi}(\vec{k}_{i}^2)\Bigg)^{"}  
\nonumber \\ 
& =  &\frac{1}{\pi\,N} \int \!\!\int d\bar{s} \; d(\frac{s_i\!-\!s_f}{Q})\; \; 
\frac{ (2\bar{s}+Q^2) \; \theta(\cdots) }{D^{3/2}}
 \;\phi(s_f)\;\phi(s_i) \times \! \int\frac{d\phi}{2}\, .
\label{ff1:pf} 
\end{eqnarray} 
It is noticed that there is no extra factor at the l.h.s. 
of eq. (\ref{eq:jac-pf}) or in the integrand entering 
the expression of the form factor in terms of the spectator momentum, 
first line of eq.  (\ref{ff1:pf}). 
As the ``point-form" form factor is Lorentz invariant, 
the extra factor required in the instant form to ensure that the
charge be Lorentz invariant is not needed here. After performing the
last integration over the $\phi$ angle, one gets a factor $\pi$, 
allowing one to recover eq. (\ref{eq:ff1-disp2}).

%
\section{Quantitative effects due to the implementation of constraints}

\begin{figure}[htb]
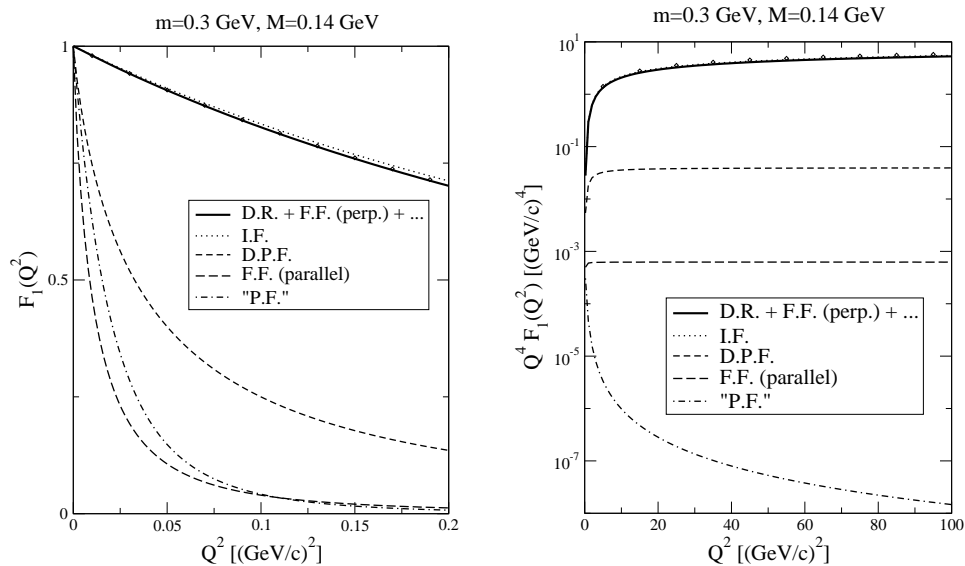

\begin{center}
\mbox{ \epsfig{ file=frat1s.eps, width=6cm}}
\hspace{3mm}
\mbox{ \epsfig{ file=frat1S.eps, width=6cm}}
\end{center}
\caption{Charge form factor in various forms of relativistic quantum 
mechanics: without and with effect of restoration of properties related 
to space-time translations,  together with currents determined in this work. 
Results at low $Q^2$ are shown in the left panel while those at high  $Q^2$, 
multiplied by the factor $Q^4$  are shown in the right panel. 
It is noticed that the thick line represents at the same time 
the dispersion-relation ones (D.R.), the front-form ones in the perpendicular 
configuration (F.F.~(perp)) and all the other results with restoration 
(dots in the inset). 
The comparison of these results with the ``exact" ones, represented by diamonds, 
evidences a very slight discrepancy. See text for other details.
\label{fig2} }
\end{figure}  

\begin{figure}[htb]
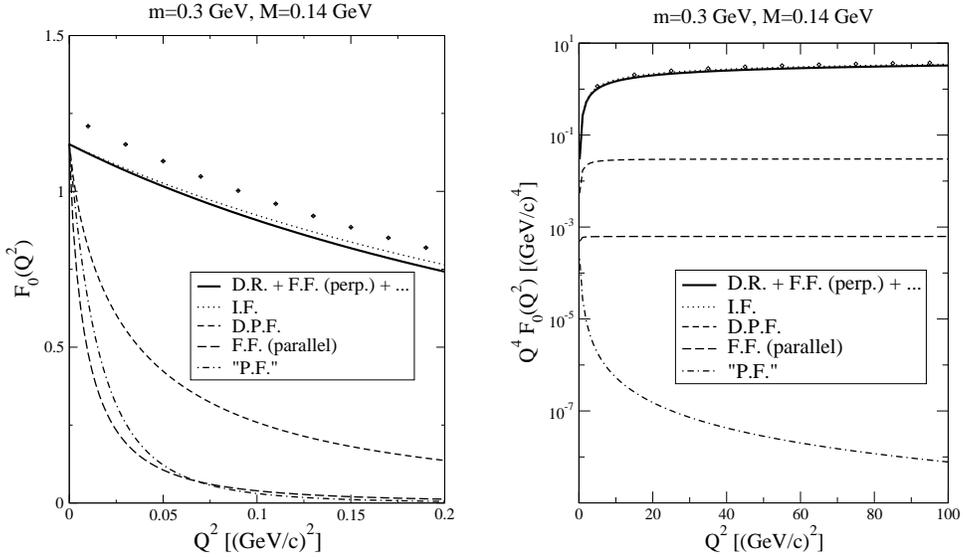

\begin{center}
\mbox{ \epsfig{ file=frat0s.eps, width=6cm}}
\hspace{3mm}
\mbox{ \epsfig{ file=frat0S.eps, width=6cm}}
\end{center}
\caption{Scalar form factor in various forms of relativistic quantum 
mechanics: without and with effect of restoration of properties related 
to space-time translations,  together with currents determined in this work. 
The comparison of these results with the ``exact" ones   
evidences that some significant discrepancy remains at low $Q^2$. 
See caption of fig. \ref{fig2} or text for other details.
\label{fig3} }
\end{figure}

\begin{figure}[htb]
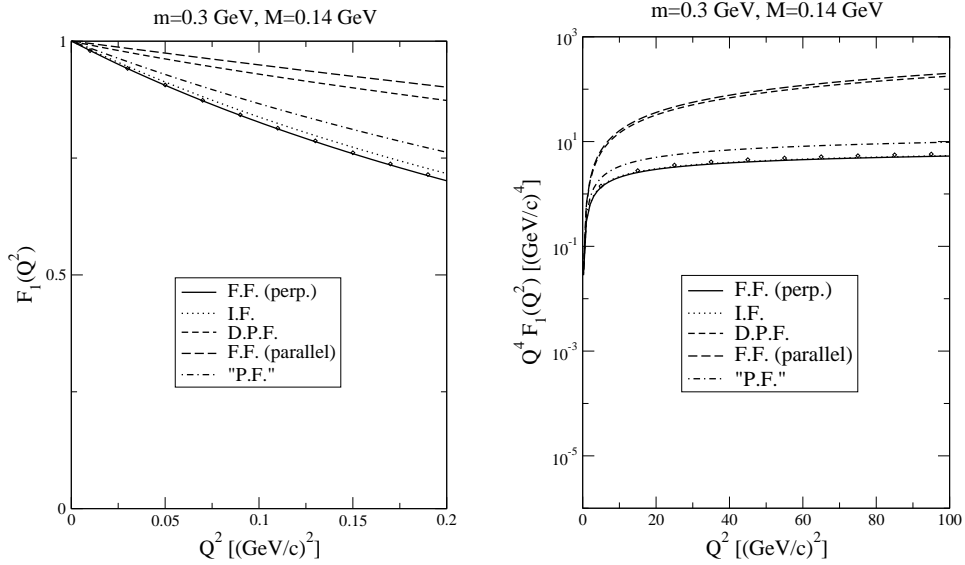

\begin{center}
\mbox{ \epsfig{ file=fraf1s.eps, width=6cm}}
\hspace{3mm}
\mbox{ \epsfig{ file=fraf1S.eps, width=6cm}}
\end{center}
\caption{Charge form factor in various forms of relativistic quantum 
mechanics: with effect of restoration of properties related 
to space-time translations together with free-particle currents, 
as most often retained. 
See caption of fig. \ref{fig2} or text for other details.
\label{fig4} }
\end{figure}  

We consider in this section some quantitative aspects 
of the restoration of properties related to space-time translations. 
This is done for the ground state of a system of scalar particles interacting 
through the exchange of a massless scalar boson 
(Wick-Cutkosky model~\cite{Wick:1954,Cutkosky:1954}). 
Exact solutions describing this system can be obtained by solving 
the Bethe-Salpeter equation, allowing one to calculate form factors.  
These ones, denoted as ``exact" ones, can be considered as a benchmark 
that RQM calculations 
described in this paper should be compared with. The wave functions 
entering these last calculations are those of a quadratic mass operator 
and the parameters correspond to a pion-like system with constituent masses 
of $m=0.3$ GeV and a total mass $M=0.14$ GeV  \cite{Desplanques:2004sp}. 
The form of the interaction fulfills minimal requirements 
that are essential in reproducing the low-energy spectrum
and the asymptotic behavior of form factors. 
It only contains an effective coupling constant
as a parameter (see eq. (56) of ref. \cite{Amghar:2002jx}).
The choice of the ingredients, $m$ and $M$, is interesting with two respects.
On the one hand, due to a large binding, we expect that the form factors 
should evidence important relativistic effects. Such a case is better 
to check that the incorporation of relativistic effects 
considered here is correctly done. Systems with a small binding considered
elsewhere \cite{Amghar:2002jx, Junhe:2004, Desplanques:2004} 
would not be so probing. Moreover, in these cases, 
discrepancies due to an incomplete account of relativistic effects 
could have a size comparable to discrepancies with an ``exact" calculation, 
due to uncertainties in determining the mass operator. 
On the other hand, the choice of the system with a total mass much smaller 
than the sum of the constituent masses has revealed a paradox. 
Getting a small total mass can be achieved by increasing the attraction 
between constituents. While it is then expected that the system shrinks, 
some approaches have instead show that the charge radius was increasing and
could go to $\infty$ while $M \rightarrow 0$ \cite{Desplanques:2004}. 
This strange result could be the sign that some important symmetries, 
like those considered in this paper, are missed. We will come back 
on the role of the binding energy at the end of this section, after having
considered effects due to properties expected from space-time translations.

Three sets of results are discussed here. The first one, which 
involves the charge form factor, $F_1(Q^2)$, is aimed to show 
that the restoration of properties related to space-time translations 
can remove large discrepancies between different RQM approaches. 
The se\-cond one is concerned with the scalar form factor, $F_0(Q^2)$. 
Its interest mainly resides in the comparison to the ``exact" results, 
especially at low $Q^2$, where its value is not constrained 
by the conservation of some charge as for $F_1(Q^2)$.
The third one provides results for $F_1(Q^2)$ with a
single-particle current that is proportional to the free-particle one, 
as employed in most RQM works. We stress that this current, 
contrary to those determined in the present work, do not necessarily 
allow one to fulfill Lorentz invariance. The three sets of results are
successively shown in figs. \ref{fig2}, \ref{fig3}, \ref{fig4}. 
In all cases, we present two panels correspon\-ding to low- and high-momentum  
transfers. The first one, for $0 \leq Q^2 \leq 0.2 \,({\rm GeV/c})^2$, 
is motivated by the presence of large effects due the smallness 
of the total mass, $M$,  in comparison to the sum 
of the constituent masses, $2m$. It is reminded 
that the slope at $Q^2=0$ is directly related to the square radius. 
The second panel, for $0 \leq Q^2 \leq 100 \,({\rm GeV/c})^2$, 
is dealing with the asymptotic behavior, 
expected to be $Q^{-4}$ up to log terms. Thus, the product 
of these last results by $Q^{4}$, actually shown in the figures, 
should evidence some plateau at high $Q^2$.

The results presented in the figures involve the dispersion-relation ones 
(D.R., eqs.~(\ref{eq:ff1-disp},~\ref{eq:ff0-disp})), 
which are frame independent, while the other ones, 
corresponding to the instant-form (I.F., eqs. (\ref{ifx1}, \ref{ifx0})), 
the front-form (F.F., eqs. (\ref{eq:rqm-ff5})), and the point-form,  
are obtained in the Breit frame. 
For the front-form case, due to the dependence on the front orientation, 
we consider the two extreme cases  of a ``perpendicular" and a ``parallel"
orientation (respectively $  \vec{q}\perp\hat{n}$ and 
$\vec{q} \parallel \hat{n}$). It is noticed that the front-form results 
for the charge form factor in a ``perpendicular"  configuration ($q^+=0$), 
without or with restoration of properties related to space-time translations,
always coincide with the dispersion-relation approach ones, 
as expected from eqs. (\ref{2q}, \ref{2r}). For the point-form case, 
we consider the Bakamjian one \cite{Bakamjian:1961} (``P.F.", eq. \ref{ff1:pf}) 
and a Dirac-inspired one  \cite{Desplanques:2004rd} (D.P.F.), 
which can be obtained by appropriately weighting the front-form results,
eqs.~(\ref{eq:rqm-ff5}), over the front orientation \cite{Desplanques:2004sp}.
Undisplayed expressions for the scalar form factor are obtained 
as for the instant-form case (see explanation before eq. (\ref{ifx0})).

Examination of both panels of fig. \ref{fig2} for the charge form factor 
shows that the restoration of properties related to space-time translations 
involves very large effects in some cases. After accounting for them, 
all results coincide with the dispersion-relation or the front-form ones 
in the ``perpendicular" configuration, as expected from
the determination of the currents. At low $Q^2$, it is noticed that the fast
drop off of the form factor appearing in some cases has disappeared. 
This result supports the suggestion made elsewhere \cite{Desplanques:2004} 
that this strange behavior of the form factor was in relation 
with some symmetry breaking. 
At high $Q^2$, the right power-law behavior of the charge form factor 
is obtained after restoring  properties related to space-time translations, 
though it was not always so before (``P.F."). We notice that the way the
equivalence of various approaches is obtained is not trivial. 
The integrands entering the integration over the spectator momentum, 
$\vec{p}$, differ, sometimes by a large amount. 
The remaining discrepancy with ``exact" results is rather small: 
0.4\% at $Q^2=0.2\,({\rm GeV/c})^2$, 12\% at $Q^2=100\,({\rm GeV/c})^2$. 
At low $Q^2$, this discrepancy is limited by the requirement that the
charge should be recovered. At high $Q^2$, the absence of a large discrepancy
points, first, to the adequacy of the overall choice of the mass operator 
in determining the asymptotic behavior of the  wave function at large momentum 
and, secondly, to the accurateness of the parameters entering 
its derivation. We cannot discard the necessity to introduce 
further two-body currents but there is no strong requirement for them.

Examination of both panels of fig. \ref{fig3} for the scalar form factor
evidences features very similar to those observed for the charge form factor. 
The main difference concerns the comparison with ``exact" results at low $Q^2$.
Contrary to the last one, the scalar form factor at $Q^2=0$ is not protected 
by some conservation law like the charge-current one. 
While results for the charge form factor were not requiring 
further significant contribution, the present ones do require some. 
The remaining discrepancy, which concerns all approaches, points 
to two-body currents that are different from those implicitly considered here
and were concerned with the restoration of some symmetry properties.
The requirement for such extra currents is thus more obvious here 
than for the charge form factor. For simplicity, we have not presented here
results for a model corresponding to the exchange of an infinitely-heavy boson, 
also described by a standard triangle Feynman diagram
\cite{Desplanques:2004sp}. In this case, the description 
of either form factor does not necessitate any extra current. 
The required two-body currents are therefore expected to imply 
the dynamics of the system under consideration 
rather than the implementation of relativity.

In fig. \ref{fig4}, we show results for the charge form factor 
obtained with a current proportional to the single-particle contribution, 
$\propto (p_i+p_f)^{\mu}$. This choice is most often assumed 
for scalar constituents in RQM approaches. 
The current also contains factors that ensure minimal
properties like making the charge independent of the momentum 
of the system. Its form has been given in ref. \cite{Desplanques:2004sp}.
Results without restoration of properties related to space-time translations, 
not displayed here, are close to those shown in fig. \ref{fig2} 
(within a few 10\% for the left panel and a factor 3 at most 
for the right one). Results with restoration were presented 
in ref. \cite{Desplanques:2004sp} using an analytical method 
when possible (I.F. (Breit frame), ``P.F."), and a numerical one 
in the other cases (F.F.~(parallel), D.P.F.). 
From the first cases, there was no indication 
that the method could have some relevance. Results presented here 
are all based on the analytical method developed in this work 
and those for the F.F.~(parallel) and D.P.F. cases are therefore new ones.
At low $Q^2$, the fast drop off of form factors in some cases has disappeared, 
supporting the suggestion that this feature was due to missing important
symmetry properties. The remaining spreading of the results 
can be ascribed for a part to the chosen current, independently of the fact 
that  Lorentz invariance of the form factor is violated (F.F. and I.F. cases)
or preserved (D.P.F. case). 
This is better seen on the high $Q^2$  part of fig.  \ref{fig4} where results 
for the F.F.~(parallel) and D.P.F. cases tend to depart from the other ones.
The analysis of the F.F.~(parallel) form factor, which can be identified 
to an instant-form one in an infinite-momentum frame, rather evidences 
a $Q^{-2}$ power-law behavior while the instant-form form factor 
in the Breit frame is consistent with the expected $Q^{-4}$ power-law 
of the ``exact" result. The discrepancy should increase with $Q^2$. 
Within the $Q^2$ range considered here and for the strongly bound system 
of interest in this work, the discrepancy is however much smaller than 
before restoring properties related to space-time translations. 
All drawbacks related to the smallness of the total mass of the system 
with respect to the sum of the constituent ones have disappeared.

For the purpose of illustrating the role of properties related 
to space-time translations, we looked at a system with a strong binding. 
At first sight, one expects that effects should be smaller in a system 
with a small binding (like the deuteron). This is true in many cases 
but not all however. A first exception concerns the ``point-form" approach.  
It evidences a change in the power-law asymptotic behavior 
of the uncorrected form factors, which shows up  as soon as 
the momentum transfer is large enough (a few constituent masses, 
see ref. \cite{Allen:2000ge} about the deuteron case 
and refs. \cite{Amghar:2002jx,Desplanques:2004} for a system
similar to the one considered in this work).
The discrepancy with results from other approaches could then compare 
to the discrepancy with experiment. Results corrected for properties 
related to space-time translations, by eliminating discrepancies due to an
incomplete implementation of relativity, can therefore give a better insight 
on the role of the dynamics in explaining other discrepancies.
A second exception concerns current hadronic systems.
As already mentioned, the discrepancy between different approaches 
can be ascribed to the appearance of a factor $4e_k^2/M^2$ 
multiplying the square momentum transfer $Q^2$ in some cases.  
Contrary to the pion-like system considered in this section, 
the apparent binding energy could be relatively small 
($\rho$ meson for instance). Due however to a confining potential, 
the momentum carried by the constituents can be quite large 
and the above factor could depart from the value 1 to reach values 
as large as 2. This is not as much as for the pion-like system we considered 
(a factor 30) but it can nevertheless produce large effects. 
Again, this factor is removed by accounting for properties related 
to space-time translations with the consequence that one can better disentangle 
in some process the role of the hadronic dynamics, which we would like 
to learn about, and of symmetries that are  implied by Poincar\'e covariance.
An interesting extension of the above discussion concerns the nucleon form
factors, where the quark mass is often taken as close to $1/3$ 
of the nucleon mass. Some of the discrepancies 
between the ``point-form" results \cite{Wagenbrunn:2001}
and other ones can be traced back to the factor that generalizes 
the above one,  $4e_k^2/M^2$, to the three-body case 
and can be of the order of 3. 
Accounting for  properties related to space-time translations 
would remove the nice agreement with experiment obtained by the authors 
but, then, there would be some role for an important piece of physics 
they ignored and is known under the name of ``vector-meson dominance". 
\section{Conclusion}
In this paper, we examined the consequences of Poincar\'e space-time
translation invariance for the calculation of form factors in RQM approaches.
This symmetry implies energy-momentum conservation but, most of all, 
it supposes that the current describing the interaction with the external probe 
transforms covariantly under the above translations. 
In practice, energy-momentum conservation is of course assumed 
in RQM calculations but, in absence of sensible tests, conditions 
to obtain this result are not generally checked. This contrasts 
with symmetries related to other transformations like rotations 
or boosts, which can be easily verified. 

We considered the problem and looked at relations involving 
the commutators of the current with the 4-momentum operator, 
$P^{\mu}$, that generate space-time translations. Ge\-ne\-rally, 
these relations turn out to be violated with a one-body-like current. 
In a particular case, they  
point out to the fact that the square momentum transferred to the
constituents differs from that one  transferred to the system.
One solution to enforce these last relations has been elaborated, 
which actually amounts to add selected contributions of many-body currents 
at all orders in the interaction strength. It has been applied 
to the calculation of form factors for the simplest two-body system. 
This was done in various forms of relativistic quantum mechanics,  
for the charge form factor mainly and, to some extent, 
for the scalar form factor. The procedure supposes to start 
from an appropriate one-body current in each case. Simultaneously, results 
for a dispersion-relation based approach, form independent, 
were considered.

For the charge form factor, we found that all approaches we considered 
could provide the same final expression after accounting for the covariance 
property of transformations of currents under space-time translations. 
Moreover, they turn out to agree with the dispersion-relation approach.
To reach this result however, quite different methods had to be used, 
depending on the symmetries evidenced or violated by each approach. 
In cases with axial symmetry (instant form and ``point form" 
in the Breit frame), expressions we started from for form factors 
are given by two-dimensional integrals. In all the other cases, 
for arbitrary momentum configuration or front orientation, 
these expressions are three-dimensional ones, to be reduced 
to two-dimensional integrals ultimately. Furthermore, the invariances 
of form factors under rotations and boosts, which are fulfilled 
in the ``point form", are not a priori satisfied by the front form 
for the former and by the instant form for the latter. 
An important ingredient in getting the above result concerns 
the expression of the original current. This one could depend 
on the approach but it appears to be rather simple in all cases, 
though it is not that one expected from the interaction 
of the external probe with a single-particle current proportional 
to the sum of the on-mass-shell momenta of the constituents, 
$(p_i+p_f)^{\mu}$. We stress that requiring invariances 
of form factors under boosts or rotations is essential to determine 
the current in some cases (instant and front forms respectively).
Thus, in one way or another, all aspects of transformations 
under the Poincar\'e group: boosts, 
rotations and space-time translations,  are involved in getting 
predictions from different RQM approaches converging to a common result, 
which is given by the dispersion-relation approach.  
The exact theoretical result is reproduced for a model 
with an infinite-mass exchange interaction. We however want to stress 
that one of the RQM approaches deserves a special mention. 
Form factors in the front-form with $q \ccdot \omega=0$ 
(often referred to as $q^+=0$) are unaffected by the above considerations, 
thus providing support to an approach which is generally believed 
to be more reliable than other ones.

Many of the above remarks hold for the scalar form factor. 
The main difference concerns the current to start with in various forms. 
While it is always possible to find such a current, the form we got for it 
from the most straightforward derivation is not simple and does not seem 
to have much theoretical support. This is probably the indication 
that the implementation of constraints related to space-time 
translations could be more sophisticated than what we assumed. 
This contrasts with the charge form factor. 
A reason could be that, in this case, the charge is closely related to the
orthogonalization of different states and the existence of an underlying 
conserved current.  

The effect of the restoration of properties related to space-time 
translations has been considered in the case of a strongly bound system 
for which a theoretical model is available. Large discrepancies between
different approaches, for both the charge or scalar radius 
and the asymptotic behavior could thus be completely removed.
All discrepancies related to the smallness of the total mass (in comparison to
the sum of the constituents masses) have disappeared. The charge form factor so
obtained is relatively close to the theoretical one over the whole range
considered here. The comparison for the scalar form factor shows similar
features at high $Q^2$. At low  $Q^2$ however, some significant discrepancy
remains (10\% at $Q^2=0$), indicating that the corrections we considered do not
exhaust all contributions due to two-body currents. 
We also looked at charge form factors with a current proportional to the free
particle one, $(p_i+p_f)^{\mu}$. This one does not guarantee that all 
properties expected from relativity are fulfilled. While effects due to the
strong binding disappeared with the implementation of properties related to
space-time translations, sizable discrepancy between some approaches, 
involving a change in the asymptotic power-law behavior of the form factor, 
are still present. This shows the importance of an appropriate determination 
of the current in getting correct results in some cases.

We believe that the present work shows unambiguously that constraints 
related to the covariant transformations of currents under space-time 
translations have an important role in providing reliable estimates 
of form factors. Having a somewhat geometrical cha\-rac\-ter, in relation 
with the choice of the hypersurface on which physics is described, 
it is not surprising that these constraints, together with properties 
from other transformations generated by the Poincar\'e-group generators 
(boosts and rotations), provide the same results for form factors. 
This result realizes the expectation that different RQM approaches should be
equivalent up to a unitary transformation \cite{Sokolov:1978}. 
The constraints we considered amount to account for many-body currents, 
not all however as the example of the scalar form factor shows, 
pointing out to further studies. Moreover, in this work, we only look 
at the constraint from a double commutator involving the variable $q^2$, 
for which a solution could be found. This was sufficient for our purpose 
as form factors considered here are exclusively depending on this variable.
More general solutions, involving for instance the $q^{\mu}$ variable 
besides $q^2$, or its powers, are not excluded however.
It is not clear whether accounting for these extra constraints 
could still be tractable. In any case, the conside\-ra\-tion of these extra
constraints could be required in the case of more complex systems 
with a non-zero spin. Further studies would be useful here too.

This work is partly supported by the National Sciences Foundations of China
under grant No. 10775148.
Y.B. D. would like to thank LPSC (Grenoble) for its hospitality.

%
\appendix

\section{ Derivation of $I(s_i,Q^2,s_f)$ in the spinless case for an arbitrary
momentum configuration}
\label{app:fnI}
We give below details about the demonstration of eq. (\ref{eq:fnI}).
We remind that $\tilde{P}_i^2=s_i$, $\tilde{P}_f^2=s_f$, 
$(\tilde{P}_i\!-\!\tilde{P}_f)^2=q^2=-Q^2$ and that the 4-vectors 
$\tilde{P}^{\mu}$ do not verify the usual on-mass-shell conditions 
($\tilde{P}^2 \neq M^2$). Noticing that 
$\int \frac{d\vec{p}_{i,f}}{2e_{i,f}}=\int d^4p_{i,f}\;\delta (p_{i,f}^2-m^2)$ 
(positive $p_0$), and $p_{i,f}^{\mu}=\tilde{P}_{i,f}^{\mu}\!-\!p^{\mu}$ 
(from the $\delta^4(\cdots)$ functions), 
eq. (\ref{eq:def-fnI}) defining $I(s_i,Q^2,s_f)$   writes:
\begin{eqnarray}
&&I(s_i,Q^2,s_f)= \frac{1}{2\pi}\int \frac{d\vec{p}}{e_p}\;
 \delta(s_i-2 p \ccdot \tilde{P}_i)  \; 
 \delta(s_f-2 p \ccdot \tilde{P}_f ) \, .
\label{eq:fnI1}
\end{eqnarray}
In order to make the integration over $\vec{p}$, we assume, 
without lost of generality, that the momenta, 
$\vec{\tilde{P}}_i, \;\vec{\tilde{P}}_f$, are in the $x,\,y$ plane.
As suggested by the above equation, we express the components of $\vec{p}$ 
in terms of $p \ccdot \tilde{P}_i $, $p \ccdot \tilde{P}_f $ and $p^z$.  
We thus obtain: 
\begin{eqnarray}
&&p^x =\frac {e_p(\tilde{P}_i^0\tilde{P}_f^y\!-\!\tilde{P}_f^0\tilde{P}_i^y)
-\frac{1}{2}\Big(s_i\tilde{P}_f^y \!-\!s_f\tilde{P}_i^y \Big)
}{\tilde{P}_i^x\tilde{P}_f^y\!-\!\tilde{P}_i^y\tilde{P}_f^x } \, ,
\nonumber \\&&
p^y =\frac {e_p(\tilde{P}_i^0\tilde{P}_f^x\!-\!\tilde{P}_f^0\tilde{P}_i^x)
-\frac{1}{2}\Big(s_i\tilde{P}_f^x \!-\!s_f\tilde{P}_i^x \Big)
}{\tilde{P}_i^y\tilde{P}_f^x\!-\!\tilde{P}_i^x\tilde{P}_f^y } \, , 
\label{eq:fnI2}
\end{eqnarray}
where $e_p$, which is solution of a second-order equation, is given by: 
\begin{eqnarray}
e_p\!&\!=\!&\!\frac{1}{Q^2D}
\Bigg ((2\bar{s}\!+\!Q^2) (\tilde{P}_{i0}s_f\!+\! \tilde{P}_{f0}s_i) 
-2s_is_f(\tilde{P}_{i0}\!+\! \tilde{P}_{f0})
\nonumber \\ 
&&\hspace*{1.2cm} 
\pm 2Q \sqrt{ s_is_f\!-\! (m^2\!+\!p^{z2})D}
\;\sqrt {\tilde{P}_{i0}\tilde{P}_{f0}\,(2\bar{s}\!+\!Q^2) 
\!-\! (\tilde{P}_{i0}^2s_f\!+\! {\tilde{P}_{f0}}^2s_i ) 
\!-\!\frac {Q^2D}{4}} \;\Bigg).\hspace*{0.2cm} 
\label{eq:fnI3}
\end{eqnarray}
The integration volume transforms as follows:
\begin{eqnarray}
\frac{d\vec{p}}{e_p}=\sum 
\frac{2\,d(p \ccdot \tilde{P}_i )\,d(p \ccdot \tilde{P}_f )\,dp^z }{   
   Q\sqrt{ s_is_f-(m^2\!+\!p^{z2})D}}\,,\label{eq:fnI4}
\end{eqnarray}
where all the dependence on the components of the 4-vectors
$\tilde{P}^{\mu}_{i,f}$ is found to be absorbed 
into the quantities $s_i,\,s_f,\,Q^2$. After inserting 
the last result into eq. (\ref{eq:fnI1}) and integrating 
over the variables $p \ccdot \tilde{P}_i $,  $p \ccdot \tilde{P}_f $, 
using the $\delta(\cdots)$ functions,
taking also into account that there are two solutions for $e_p$, 
one gets the desired result:
\begin{eqnarray}
I(\cdots)&\!=\!&\frac{1}{8\pi}\int dp^z  \sum
\frac{2}{Q\sqrt{ s_is_f-    (m^2\!+\!p^{z2})D}}
=\frac{\theta(\cdots)}{2Q\sqrt{D}}\,.\label{eq:fnI5}
\end{eqnarray}
It is interesting to notice how the extra factor $\pi$ is obtained 
at the r.h.s.. 
In a system with azimuthal symmetry, a factor  proportional to $\pi$ simply 
arises from the integration over the azimuthal angle  $\phi$.
In the present case, where this symmetry is not assumed, it comes from an
integral of the type $\int^{|a|}_{-|a|}\frac{dz}{(a^2-z^2)^{1/2}}=\pi$.

\section{Details relative to the implementation of constraints from
transformation of currents under space-time translations}
\label{app:detail}
For a RQM description of a two-body system on some hyperplane with orientation
$\xi^{\mu}$, the momenta of the  interacting constituent, 
$p_i^{\mu}$ (or $p_f^{\mu}$), the momentum of the spectator constituent, 
$p^{\mu}$, and the total momentum, $P_i^{\mu}$ (or $P_f^{\mu}$), verify the
relation:
\begin{equation} 
\vec{p}_i+\vec{p}-\vec{P}_i=(e_i+e_p-E_i)\;\frac{\vec{\xi}_i}{\xi_i^0}\, ,
\end{equation}
and a similar one for the final state. This relation can be used 
to get eqs. (\ref{sump1p2}, \ref{Delta}), which are symmetrical 
in the momentum of the two constituents. It can also be used to express 
the momentum of the interacting constituent in terms 
of the spectator momentum, the total momentum and the hyperplane orientation:
\begin{eqnarray} 
p_i^{\mu}&\!=\!&P_i^{\mu}-p^{\mu}
+\frac {(m^2\!-\!(P_i\!-\!p)^2)\,\xi_i^{\mu}}{ (P_i\!-\!p) \ccdot \xi_i+
\sqrt{ (m^2\!-\!(P_i\!-\!p)^2)\,\xi_i^2 
+\Big ( (P_i\!-\!p) \ccdot \xi_i    \Big)^2}}   \, ,
\end{eqnarray}
which holds for a finite as well as a zero value of $\xi_i^2$.
This last relation can be employed either to calculate the square momentum 
transferred to the interacting constituent, $(p_i\!-\!p_f)^2$, 
considered in the main text, or to calculate the quantity 
$s_i^0 =(p_i\!+\!p)^2$ (or $s_f^0 =(p_f\!+\!p)^2$) which, in particular, 
enter the argument of wave functions. In the case of constituents 
with the same mass, this quantity (and a similar one for the final state) 
can be expressed as:
\begin{eqnarray} 
s_i^0 =2p \!\cdot\! P_i
+2\,\frac{(2p \!\cdot\! P_i \!-\! P_i^2)\, p \ccdot  \xi_i   }{
(P_i\!-\!p) \ccdot \xi_i+
\sqrt{ (m^2\!-\!(P_i\!-\!p)^2)\,\xi_i^2 
\!+\! \Big ((P_i\!-\!p) \ccdot \xi_i  \Big)^2}}\, .
\end{eqnarray}
The expressions of $s_i^0$ and $s_f^0$ are considered below together 
with their modification implied by constraints under discussion in this paper. 
For practical purposes, we split the total momenta of the initial 
and final states into an average one 
$\bar{P}^{\mu}=\frac{1}{2}( P_i^{\mu}\! +\! P_f^{\mu})$ 
and a difference $q^{\mu}=( P_f^{\mu}\! -\! P_i^{\mu})$
(equivalently $P_i^{\mu}=\bar{P}^{\mu}\!-\!\frac{1}{2}q^{\mu}$, 
$P_f^{\mu}=\bar{P}^{\mu}\!+\!\frac{1}{2}q^{\mu}$). At first sight, 
this  is not simplifying the consideration of the square-root terms 
which appear in the above expressions and involve contributions
of the type $\pm p \!\cdot\! q $, $\pm \bar{P} \!\cdot\! q $ 
or $\pm   q \ccdot  \xi$. However, it will be seen that, 
in the cases where the $\pm$ sign under the square-root symbol cannot be 
factored out in one way or another (instant form and generalized one), 
accounting for the above constraints allows one to do it, 
greatly contributing to make algebraic calculations tractable.

\subsection{Instant form}
\label{detail:if}
For the instant form, one has $\xi_i^{\mu}=\xi_f^{\mu}=\lambda^{\mu}$ with 
$\lambda^{\mu}=(1,\,0,\,0,\,0)$. The quantity $s_{i,f}^0$ then writes:
\begin{eqnarray}
s_{i,f}^0=2\,m^2+2\,e_p \,e_{\bar{P}\!-\!p\!\mp\!\frac{q}{2}} 
- 2 \, \vec{p}\cdot (\vec{\bar{P}}\!-\!\vec{p} \mp\!\frac{\vec{q}}{2})\,,
\label{eq:s0if}
\end{eqnarray}
where: 
\begin{eqnarray}
e_{\bar{P}\!-\!p\!\mp\!\frac{q}{2}}= 
 \sqrt{m^2+(\vec{\bar{P}}\!-\!\vec{p} \mp\!\frac{\vec{q}}{2})^2}\,.
\end{eqnarray}
Considering with some detail the square of this quantity together 
with the effect of constraints given by eq. (\ref{alpif}), 
one gets for the transformed quantity denoted with $``\cdots"$: 
\begin{eqnarray}
``e_{\bar{P}-p \mp \frac{q}{2}}^2"
\!&\!=\!&\!
``\Big (m^2+(\vec{\bar{P}}\!-\!\vec{p}\mp\!\frac{\vec{q}}{2})^2 \Big) "
= m^2+(\vec{\bar{P}}\!-\!\vec{p}\mp\!\frac{\alpha\vec{q}}{2})^2 
\nonumber \\ 
\!&\!=\!&\!m^2\!+\! (\vec{p}\!-\!\vec{\bar{P}})^2 \pm
(\vec{p}\!-\!\vec{\bar{P}})\!\cdot\!\hat{q}\sqrt{\!-q^2}\, 
\sqrt {1+\frac {((\vec{p}\!-\!\vec{\bar{P}}) \ccdot \hat{q})^2 }{ 
    m^2\!+\!(\vec{p}\!-\!\vec{\bar{P}})^2 \!
-\!((\vec{p}\!-\!\vec{\bar{P}}) \ccdot \hat{q})^2\!-\!\frac{q^2}{4} }  }
\nonumber \\ && \hspace{2cm}
-\frac{q^2}{4}
\Bigg (1+\frac{((\vec{p}\!-\!\vec{\bar{P}}) \ccdot \hat{q})^2}{m^2\!+\!
(\vec{p}\!-\!\vec{\bar{P}})^2 \!-\!((\vec{p}\!-\!\vec{\bar{P}}) \ccdot \hat{q})^2
\!-\!\frac{q^2}{4}} \Bigg)
\nonumber \\ 
\!&\!=\!&\!\Bigg (\sqrt{m^2\!+\!(\vec{p}\!-\!\vec{\bar{P}})^2\!-\!\frac{q^2}{4}}
\pm \frac {(\vec{p}\!-\!\vec{\bar{P}}) \ccdot \hat{q}\sqrt{\!-q^2}  }{
2\,\sqrt {m^2\!+\!(\vec{p}\!-\!\vec{\bar{P}})^2 \!-\!
((\vec{p}\!-\!\vec{\bar{P}}) \ccdot \hat{q})^2\!-\!\frac{q^2}{4}}  }
\Bigg)^2\, ,
\end{eqnarray} 
where the last line has the desired property:
\begin{eqnarray}
&&``e_{\bar{P}-p \mp \frac{q}{2}}"=
\sqrt{m^2\!+\!(\vec{p}\!-\!\vec{\bar{P}})^2\!-\!\frac{q^2}{4}}
\pm \frac {(\vec{p}\!-\!\vec{\bar{P}})\!\cdot\!\hat{q}\sqrt{\!-q^2}}{
2\,\sqrt{m^2\!+\!(\vec{p}\!-\!\vec{\bar{P}})^2 \!-\!
((\vec{p}\!-\!\vec{\bar{P}})\!\cdot\!\hat{q})^2\!-\!\frac{q^2}{4}}}\, .
\end{eqnarray} 
Inserting this result into eq. (\ref{eq:s0if}) and taking into account  
that the last term in this equation has also to be transformed, 
one gets for $s_i,s_f$:
\begin{eqnarray}
s_{i,f} &=& ``s^0_{i,f}"=
2\,m^2\!+\!2\,e_p \,e_{\bar{P}\!-\!p\!\mp\!\frac{\alpha q}{2}} \!-\! 
2 \, \vec{p}\ccdot (\vec{\bar{P}}\!-\!\vec{p} \mp\!\frac{\alpha \vec{q}}{2})
\nonumber \\
 &=&2 \Bigg ( m^2\!+\!\vec{p} \ccdot (\vec{p}\!-\!\vec{\bar{P}})
+\sqrt{ m^2\!+\! \vec{p}\,^2}\,
\sqrt{ m^2\!+\! (\vec{p}\!-\!\vec{\bar{P}})^2\!-\!\frac{q^2}{4}}\Bigg)
\nonumber \\
 && \pm \,
\frac {\sqrt{\!-q^2}\; \Big ( \,(\vec{p}\!-\!\vec{\bar{P}}) \ccdot \hat{q}\,
 \sqrt{ m^2\!+\! \vec{p}\,^2} + \vec{p} \ccdot \hat{q}\,
\sqrt { m^2\!+\!(\vec{p}\!-\!\vec{\bar{P}})^2\!-\!\frac {q^2}{4} } \;\Big )
  }{  \sqrt { m^2\!+\!(\vec{p}\!-\!\vec{\bar{P}})^2 \!-\!
    ((\vec{p}\!-\!\vec{\bar{P}}) \ccdot \hat{q})^2\!-\!\frac{q^2}{4}  }  }
\, . \label{ek2if4}
\end{eqnarray}
The above expression simplifies in the Breit frame where $\vec{\bar{P}}=0$ 
and $q^0=0$ ($-q^2=\vec{q}\,^2$). It then reads:
\begin{eqnarray}
(s_{i,f})_{B.f.} = 2\Bigg ( 
\sqrt{ m^2\!+\! \vec{p}\,^2} \!+\! 
    \sqrt { m^2\!+\!\vec{p}\,^2\!+\!\frac {\vec{q}\,^2}{4} } \Bigg ) 
\Bigg (\sqrt{ m^2\!+\! \vec{p}\,^2}\pm  
\frac {\vec{p} \ccdot \vec{q}   }{  2\sqrt { m^2\!+\!\vec{p}\,^2 \!-\!
    (\vec{p} \ccdot \hat{q})^2\!+\!\frac{\vec{q}^2}{4}  }  }\Bigg)\,.
\end{eqnarray}
%

\subsection{Front form}
\label{detail:ff}
The front form is characterized by the relation 
$\xi_i^{\mu}=\xi_f^{\mu}=\omega^{\mu}$ 
with $\omega^2=0$. The quantities $s_{i,f}^0$ are then given by:  
\begin{eqnarray} 
s_{i,f}^0 \!&\!=\!&\!   
 2\frac{P_{i,f}\!\cdot\!\omega}{(P_{i,f}\!-\!p)\!\cdot\! \omega}
 \; p \ccdot P_{i,f}
-\frac{p \ccdot \omega    }{(P_{i,f}\!-\!p) \ccdot \omega}\; P_{i,f}^2 
\nonumber \\
 \!&\!=\!&\! 2p \!\cdot\!(\bar{P}\mp\frac{q}{2})
+\frac{p \!\cdot\! \omega    }{
(\bar{P}\mp\frac{q}{2}\!-\!p) \!\cdot\! \omega}\; 
\Big(2p \ccdot (\bar{P}\mp\frac{q}{2}) \!-\! (\bar{P}\mp\frac{q}{2})^2\Big) \, .
\end{eqnarray}
 The transformed expression of $s_{i,f}^0$ reads:
\begin{eqnarray} 
s_{i,f} =``s_{i,f}^0"=2p \!\cdot\!(\bar{P}\mp\frac{\alpha q}{2})
+\frac{p  \ccdot  \omega    }{
(\bar{P}\mp\frac{\alpha q}{2}\!-\!p)  \ccdot  \omega}\; 
\Big(2p \ccdot (\bar{P}\mp\frac{\alpha q}{2}) 
\!-\! (\bar{P}\mp\frac{\alpha q}{2})^2\Big) \, ,
\end{eqnarray}
which, using the expression of $\alpha$ given by eq. (\ref{alpff}), 
also reads after some algebra: 
\begin{eqnarray} 
s_{i,f} \!&\!=\!&\! 2\frac{\bar{P}\!\cdot\!\omega}{(\bar{P}\!-\!p)\!\cdot\! \omega} 
\; p\!\cdot\!\bar{P}
-\frac{p\!\cdot\!\omega}{(\bar{P}\!-\!p)\!\cdot\! \omega}
(\bar{P}^2\!+\!\frac{q^2}{4})
\nonumber \\
&&\pm
\frac {- \frac{\bar{P}\cdot\omega}{(\bar{P}\!-\!p)\cdot \omega}
p\!\cdot\!q
+\frac{p \cdot \omega}{(\bar{P}\!-\!p)\cdot \omega}
\Big (\bar{P}\!\cdot\!q
-\frac{1}{2}\frac{q\cdot\omega}{(\bar{P}\!-\!p)\cdot \omega}
(\bar{P}^2\!-\!2p\!\cdot\!\bar{P}\!+\!\frac{q^2}{4})\Big)
}{ \sqrt {1 
-2 \frac{q\cdot\omega}{(\bar{P}\!-\!p)\cdot \omega}
\frac{(\bar{P}\!-\!p)\cdot q }{q^2}+
\frac{(q \cdot \omega)^2}{((\bar{P}\!-\!p) \cdot \omega)^2}
\frac{\bar{P}^2\!-\!2p \cdot \bar{P}\!+\!\frac{q^2}{4}}{q^2}} }\,.
\end{eqnarray}
For an elastic transition, like that one considered in this work, 
the above expression slightly simplifies as one can use the relations 
$\bar{P}\!\cdot\!q=0$ and $\bar{P}^2\!+\!\frac{q^2}{4}=M^2$. Moreover, 
in the Breit frame, one can use the relations 
$\vec{\bar{P}}=0,\, \bar{P}^0=\bar{E}=\sqrt{M^2\!-\!\frac{q^2}{4}},\, q^0=0 $. 
The expression can then be written in terms of the unit vector 
$\hat{n}=\vec{\omega}/\omega^0$:
\begin{eqnarray}
(s_{i,f})_{B.f.}  \!&\!=\!&\!  
2\frac{\bar{E}}{\bar{E}\!-\!e_p\!+\!\vec{p} \ccdot \hat{n}} 
\; e_p\bar{E}
-\frac { e_p \!-\! \vec{p} \ccdot \hat{n}  }{ 
\bar{E} \!-\! e_p \!+\! \vec{p} \ccdot \hat{n}  }\; M^2
\nonumber \\
&&\pm\frac { \frac{\bar{E}}{\bar{E}\!-\!e_p\!+\!\vec{p} \cdot \hat{n}}\; 
\vec{p} \ccdot \vec{q}
-\frac { e_p \!-\! \vec{p} \cdot \hat{n}  }{ 
2(\bar{E} \!-\! e_p \!+\! \vec{p} \cdot \hat{n})  }
\frac{\vec{q} \cdot \hat{n}}{\bar{E} \!-\! e_p \!+\! \vec{p} \cdot \hat{n}}\;
 (2e_p \bar{E} \!-\! M^2  )
}{ \sqrt {1 
-2 \frac{\hat{q} \cdot \hat{n}}{\bar{E} \!-\! e_p \!+\! \vec{p} \cdot \hat{n}}
\;\vec{p} \ccdot \hat{q}
+\frac{(\hat{q} \cdot \hat{n})^2  }{  
(\bar{E} \!-\! e_p \!+\! \vec{p} \cdot \hat{n})^2}\;(2e_p \bar{E} \!-\! M^2  )
 } }\,.
\end{eqnarray}
%

\subsection{``Point form"}
\label{detail:pf}
The ``point form" of interest here is characterized by the relations 
$\xi_{i,f}^{\mu}= P_{i,f}^{\mu}/\sqrt{P_{i,f}^2}$ with $M^2_i=M^2_f=M^2$ 
(elastic case). It follows the expressions for $s_{i,f}^0$:
\begin{eqnarray}
s_{i,f}^0=4\frac {(p \ccdot P_{i,f})^2}{P_{i,f}^2}=
4 \Bigg ( p \ccdot \hat{v} \sqrt{1-\frac{q^2}{4M^2}}
\mp \frac {p \ccdot q}{2M} \Bigg)^2 \, ,
\end{eqnarray}
where the unit 4-vector $\hat{v}^{\mu}$ has been defined in the main text. The
transformed expression of $s_{i,f}^0$ reads:
\begin{eqnarray}
s_{i,f} =``s_{i,f}^0"=4\frac {(p \ccdot P_{i,f})^2}{P_{i,f}^2}=
4 \Bigg ( p \ccdot \hat{v} \sqrt{1-\frac{\alpha^2q^2}{4M^2}}
\mp  \frac {\alpha \, p \ccdot q}{2M} \Bigg)^2 \, ,
\end{eqnarray}
which, using the expression of $\alpha$ given by eq. (\ref{alppf}), 
also reads: 
\begin{eqnarray}
s_{i,f} =  2\Bigg( 
\frac{\sqrt{(p \!\cdot\! \hat{v})^2 \!-\! (p \!\cdot\! \hat{q})^2\!-\!\frac{q^2}{4}}
}{\sqrt{(p \!\cdot\! \hat{v})^2 \!-\! (p \!\cdot\! \hat{q})^2}}\;
\Big((p \!\cdot\! \hat{v})^2 \!+\!(p \!\cdot\!\hat{q} )^2\Big)
\!+\!(p \!\cdot\! \hat{v})^2 \!-\!(p \!\cdot\!\hat{q} )^2  
\mp
\frac{(p \!\cdot\! \hat{v})(p \!\cdot\!q )}{
\sqrt{(p \!\cdot\! \hat{v})^2 \!-\! (p \!\cdot\! \hat{q})^2}}\Bigg)\, ,
\end{eqnarray}
where $\hat{q}^{\mu}$ has been defined in the main text. It is noticed 
that the expression does not depend explicitly on the total mass 
of the system, $M$. It simplifies in the Breit frame, where 
$\vec{\hat{v}}=0,\hat{v}^0=1,\hat{q}^0=0$, and then reads:
\begin{eqnarray}
&&(s_{i,f})_{B.f.} =2\Bigg ( \frac {
\sqrt { m^2\!+\!\vec{p}\,^2 \!-\! (\vec{p} \ccdot \hat{q})^2\!+\!\frac{\vec{q}^2}{4} } 
}{\sqrt { m^2\!+\!\vec{p}\,^2 \!-\! (\vec{p} \ccdot \hat{q})^2}  }
\Big ( m^2\!+\!\vec{p}\,^2 \!+\! (\vec{p} \ccdot \hat{q})^2 \Big )
\!+\! m^2\!+\!\vec{p}\,^2 \!-\! (\vec{p} \ccdot \hat{q})^2
\nonumber \\
&&\hspace*{5cm} \pm \frac {\vec{p} \ccdot \vec{q} \, \sqrt{m^2\!+\!\vec{p}\,^2}  }{
 \sqrt { m^2\!+\!\vec{p}\,^2 \!-\! (\vec{p} \ccdot \hat{q})^2}  }\Bigg ) \, .
\end{eqnarray}
%

\section{Generalizations to arbitrary momentum configurations or hyperplane
orientations}
\label{app:gen}
In this appendix, we consider generalizations of Breit-frame form
factors to arbitrary momentum configurations or hyperplane  orientations. 
For convenience, we now syste\-ma\-tically shift from the $q$ notation 
to  the $Q$ one for the momentum transfer together, in some cases, 
with unit  3-vectors ($\hat{q},\;\hat{q}^2=1$) or 
space-like 4-vectors ($\hat{q},\;\hat{q}^2=-1$, 
$\hat{\tilde{q}}, \; \hat{\tilde{q}}^2=-1$), 
on which final results do not depend.
The presentation follows lines adopted in the main text: 
1) give expressions for the combinations of $s_i, s_f$ with a symmetry character,
$\bar{s}$ and $s_i\!-\!s_f$,
2) invert these equations  to get components of the spectator momentum 
in terms of the above quantities and a third one to be chosen, 
3) calculate the integration volume in terms of the new variables,
4) give the expression of form factors.

\subsection{Form factors in the instant form: arbitrary momentum configuration}
\label{gen:if}
Starting from the relations, eqs. (\ref{ek2if4}), we first write the general 
expressions for $\bar{s}$ and $s_i\!-\!s_f$: 
\begin{eqnarray}
&&\bar{s} =2 \Bigg ( m^2+\vec{p}\ccdot (\vec{p}\!-\!\vec{\bar{P}})
+\sqrt{ m^2+ p^2}\,
\sqrt{ m^2+ (\vec{p}\!-\!\vec{\bar{P}})^2\!+\!\frac{1}{4}Q^2}\Bigg)\, , 
\nonumber \\
&&s_i\!-\!s_f= 2Q\;
\frac{(\vec{p}\!-\!\vec{\bar{P}}) \!\cdot\! \hat{q}\,\sqrt{ m^2\!+\! p^2}  +
\vec{p}\!\cdot\!\hat{q}\,
\sqrt{ m^2\!+\!(\vec{p}\!-\!\vec{\bar{P}})^2\!+\!\frac{1}{4}Q^2}
}{ \sqrt{ m^2\!+\!(\vec{p}\!-\!\vec{\bar{P}})^2\!-\!
((\vec{p}\!-\!\vec{\bar{P}})\!\cdot\!\hat{q})^2\!+\!\frac{1}{4}Q^2}}\;
\, , 
 \label{eq:gen-if1}
\end{eqnarray}
where we defined $\hat{q}=\vec{q}/|\vec{q}|$.
As the problem under consideration here involves two vectors, the momenta of the
initial and final momenta, it is appropriate to separate the components of the
spectator constituent into two components, $p^x$ and $p^y$,  belonging to the
plane defined by these two directions, and a last one, $p^z$, perpendicular to
this plane. We can thus use the above relations to express the quantities 
$\vec{p} \ccdot \vec{\bar{P}}$ and $\vec{p} \ccdot \hat{q}$ in terms of 
the variables $\bar{s}$,  $s_i\!-\!s_f$ and $p^z$. Actually, it is simpler 
to first express these quantities in terms of $\bar{s}$,  $s_i\!-\!s_f$ 
and $e_p$ and then express $e_p$ in terms of $\bar{s}$,  $s_i\!-\!s_f$ 
and $p^z$. We thus get:
\begin{eqnarray}
&&\vec{p} \ccdot \vec{\bar{P}}=\frac{1}{2} 
\Big ( e_p\sqrt{D0}-\bar{s} \Big)\, ,
\nonumber  \\
&&\vec{p} \ccdot \hat{q}=\vec{\bar{P}} \ccdot \hat{q}+
\frac{(\sqrt{D0}-2e_p) }{2 D1}
 \Bigg (\!\!-\!2\vec{\bar{P}}\!\cdot\!\hat{q}\sqrt{D0}
+\frac {s_i\!-\!s_f  }{Q}\sqrt{D2} \Bigg)\, ,
\label{eq:gen-if2}
\end{eqnarray}
where we introduced the notations:
\begin{eqnarray}
&&D0= 4\bar{s}\!+\! Q^2 \!+\! 4\vec{\bar{P}}^2\, ,
\nonumber \\
&&D1= 4\bar{s}\!+\! Q^2\!+\! \frac{(s_i\!-\!s_f)^2}{Q^2}
  \!+\! 4\vec{\bar{P}}^2\, ,
\nonumber \\
&&D2= 4\bar{s}\!+\! Q^2\!+\! \frac{(s_i\!-\!s_f)^2}{Q^2}
  \!+\! 4\vec{\bar{P}}^2\!-\!4(\vec{\bar{P}} \!\cdot\!\hat{q})^2\, .
 \label{eq:gen-if3}
\end{eqnarray}
The quantity $e_p$, in terms of the variables
 $\bar{s}$,  $s_i\!-\!s_f$ and $p^z$, is given by: 
\begin{eqnarray}
e_p=\frac {1}{2} \Bigg (\sqrt{D0}
- D1\frac {\sqrt{\frac{D2}{D}}\,(2\bar{s}\!+\! Q^2)
\pm  4\sqrt {\vec{\bar{P}}^2\!-\!(\vec{\bar{P}} \!\cdot\!\hat{q})^2} \; 
\sqrt {\frac{s_is_f}{D}\!-\!(m^2\!+\!p^{z2})}}{\sqrt{D} \,
\Big (\sqrt{D0\,D2} +2\frac{s_i\!-\!s_f}{Q}
\vec{\bar{P}} \cdot \hat{q} \Big)} \;\Bigg )\, , 
 \label{eq:gen-if4}
\end{eqnarray}
where $D$ has been defined in eq. (\ref{eq:theta} ). 
In the change of variables $p^x,\,p^y,\,p^z$  to the variables 
$\bar{s}$,  $s_i\!-\!s_f$ and $p^z$,  
the integration volumes transform as follows: 
\begin{eqnarray}
&& \hspace*{-0.5cm}\frac{d\vec{p}}{e_p} \; 
\frac{``(e_f+e_i+2e_p)"}{2\,``(e_f+e_i)"}
=\sum \frac{1}{4}\frac {d\bar{s} \; d(\frac{s_i\!-\!s_f}{Q})\, dp^z  }{  
\sqrt{D} \sqrt{\frac{s_is_f}{D}\!-\!(m^2\!+\!p^{z2})}} 
\nonumber \\ 
&& \hspace*{2.5cm} \times 
 \Bigg ( \frac{(2\bar{s}\!+\!Q^2)\pm 4\sqrt{\frac{D}{D2}}
 \sqrt {\vec{\bar{P}}^2\!-\!(\vec{\bar{P}} \!\cdot\!\hat{q})^2}\;
 \sqrt {\frac{s_is_f}{D}\!-\!(m^2\!+\!p^{z2})} }{D} \Bigg)  \,,
 \label{eq:gen-if5}
\end{eqnarray}
where a factor, which, in itself, violates Lorentz invariance, 
has been introduced at the l.h.s. so that the expected Lorentz invariance 
of the final result be fulfilled. This factor, that was also found 
for the Breit-frame case, ensures that the charge be Lorentz invariant. 
The $\sum$ symbol reminds that a summation should be done 
on the two solutions for $e_p$ given by eq. (\ref{eq:gen-if4}).
The final expression for the charge form factor thus reads:
\begin{eqnarray} 
``F_1(Q^2)"&\!=\!&\frac{16\pi^2}{N}\! \int\! \frac{d\vec{p}}{(2\pi)^3}\;
\frac{1}{e_p} \;
\frac{``(e_f+e_i+2e_p)"}{2\,``(e_f+e_i)"}\;
^{``}\Bigg(\tilde{\phi}(\vec{k_f}^2)\;\tilde{\phi}(\vec{k_i}^2) \Bigg)^{"}
\nonumber \\
&\!=\!&\frac{2}{\pi N}\!\int \!\! \int d\bar{s} \; d(\frac{s_i\!-\!s_f}{Q})\;
 \frac{  \theta(\cdots)  }{ 4D^{3/2}} \;\phi(s_f)\;\phi(s_i)
 \nonumber \\
&& \hspace*{0.6cm} \times \! \sum \!
\int\! dp^z   
  \frac{(2\bar{s}\!+\!Q^2)\pm 4\sqrt{\frac{D}{D2}}
\sqrt {\vec{\bar{P}}^2\!-\!(\vec{\bar{P}} \!\cdot\!\hat{q})^2}\;
\sqrt{\frac{s_is_f}{D}\!-\!(m^2\!+\!p^{z2})} }{ 
 \sqrt{ \frac{s_is_f}{D}\!-\!(m^2\!+\!p^{z2})} } \, .
\label{eq:gen-if6} 
\end{eqnarray} 
The sum over the two solutions of $e_p$ performed at the last line 
is important. Apart from a factor 2 for the term proportional 
to $2\bar{s}\!+\!Q^2$, it makes the terms dependent on the momentum 
$\vec{\bar{P}}$ to cancel, allowing one to get a Lorentz-invariant result.
After performing the integration over $p^z$, the expression 
for the form factor $F_1(Q^2)$, eq. (\ref{eq:ff1-disp2}), is easily recovered.

%
\subsection{Form factors in the front form: arbitrary momentum configuration }
\label{gen:ff}
The transformed expressions of the $s_{i,f}^0$ variables have been derived in
appendix \ref{detail:ff}. The combinations with a symmetry character 
are given by:
\begin{eqnarray}
&&\bar{s}=
2\frac{\bar{P}\!\cdot\!\omega}{(\bar{P}\!-\!p)\!\cdot\! \omega} 
\; p\!\cdot\!\bar{P}
+\frac{1}{4}\frac{p\!\cdot\!\omega}{(\bar{P}\!-\!p)\!\cdot\! \omega}
(Q^2\!-\!4\bar{P}^2)\,,
\nonumber \\
&&s_i\!-\!s_f=2Q\;
\frac {- \frac{\bar{P}\cdot\omega}{(\bar{P}\!-\!p)\cdot \omega} \,
p\!\cdot\!\hat{q}
+\frac{p \cdot \omega}{(\bar{P}\!-\!p)\cdot \omega}
\Big (\bar{P}\!\cdot\!\hat{q}
+\frac{1}{8}\frac{\hat{q}\cdot\omega}{(\bar{P}\!-\!p)\cdot \omega}
(Q^2\!-\!4\bar{P}^2\!+\!8p\!\cdot\!\bar{P})\Big)
}{ \sqrt {1 
+2 \frac {\hat{q} \cdot \omega}{(\bar{P}\!-\!p)\cdot \omega}
(\bar{P}\!-\!p) \ccdot \hat{q} +
\frac{1}{4}\frac{(\hat{q} \cdot \omega)^2}{((\bar{P}\!-\!p) \cdot \omega)^2}
(Q^2\!-\!4\bar{P}^2\!+\!8p\!\cdot\!\bar{P})} }\,,
\label{eq:gen-ff1} 
\end{eqnarray}
where $\hat{q}^{\mu}=q^{\mu}/\sqrt{-q^2}$. 
The problem under consideration in this subsection involves three different
orientations. In making the change of variables $p^x,\,p^y,\,p^z$ to the variables 
$\bar{s}$,  $s_i\!-\!s_f$, we choose to take the third one as 
$p\!\cdot\! \omega $. At first sight, there is no strong argument for this
choice but expressions so obtained evidence a simple dependence on the ratio 
$\frac{p \cdot \omega}{\bar{P}  \cdot \omega}$, which, actually, can be
identified to the $x$ variable usually introduced in the light-front approach.
The expressions of the scalar products, $ p\!\cdot\!\bar{P}$ 
and   $ p\!\cdot\!\hat{q}$ thus read:
\begin{eqnarray}
&&p\!\cdot\!\bar{P}=\frac{1}{2} \Bigg(\bar{s} - \frac{1}{4}
\frac{p\!\cdot\! \omega }{ \bar{P}\!\cdot\!\omega }
\Big(4\bar{s}\!+\!Q^2\!-\!4\bar{P}^2\Big)  \Bigg) \,,
\nonumber \\
&&p\!\cdot\!\hat{q}=\frac{p\!\cdot\! \omega }{ \bar{P}\!\cdot\!\omega }
\Bigg (\bar{P}\!\cdot\!\hat{q}  \!+\!  \frac{1}{8}
\frac {\hat{q}\!\cdot\!\omega}{\bar{P}\!\cdot\! \omega}
\Big (4\bar{s}\!+\!Q^2\!-\!4\bar{P}^2
\Big) \Bigg)
-\frac {s_i\!-\!s_f}{2Q}
\Big(\sqrt {E2}\!+\! \frac{\hat{q}\!\cdot\! \omega }{ \bar{P}\!\cdot\!\omega }\,
\frac {s_i\!-\!s_f}{2Q}\Big)
\Big (1\!-\!\frac{p\!\cdot\! \omega }{ \bar{P}\!\cdot\!\omega } \Big)\,,
\nonumber \\
&& \hspace*{-0.5cm}{\rm with} \hspace*{0.5cm}
E2=1
\!+\!2\frac{\hat{q}\!\cdot\! \omega }{ \bar{P}\!\cdot\!\omega } 
\bar{P}\!\cdot\!\hat{q}
\!+\!\frac{1}{4}
\Big (\frac {\hat{q}\!\cdot\! \omega }{ \bar{P}\!\cdot\!\omega }\Big)^2
\Big (4\bar{s}\!+\!Q^2\!-\!4\bar{P}^2
\!+\!\frac{(s_i\!-\!s_f)^2}{Q^2}\Big)\,.
\label{eq:gen-ff2} 
\end{eqnarray}
In the change of variables, the integration volume transforms as follows:
\begin{eqnarray}
&& \hspace*{-0.5cm}\frac{d\vec{p}}{e_p}  
\frac {(``p_i\!+\!p_f\!+\!2p)" \ccdot \omega}{2``(p_i\!+\!p_f)" \ccdot \omega}
=\frac{d\vec{p}}{e_p} \; 
\frac {\bar{P} \ccdot \omega}{2(\bar{P}\!-\!p) \ccdot \omega} \;
\nonumber \\ 
&& \hspace*{0.5cm}=\sum \frac{1}{4}\frac { d\bar{s} \; d(\frac{s_i\!-\!s_f}{Q})\, d(\frac{p \cdot \omega}{\bar{P} \cdot \omega })  }{  
\sqrt{D}\sqrt {\Big  (\frac {s_i\,s_f}{D }\!-\!m^2\Big )f
-(\frac{p \cdot \omega}{\bar{P} \cdot \omega}\!-\!d)^2} }
 \Bigg ( \frac{2\bar{s}\!+\!Q^2 }{ D}
 -(\frac{p \ccdot \omega}{\bar{P} \ccdot \omega}\!-\!d)\,g \Bigg) 
 \,, \label{eq:gen-ff3} 
\end{eqnarray}
where: 
\begin{eqnarray}
&&d=1-\frac { (2\bar{s}\!+\!Q^2) \sqrt {E2} }{
D \Big (\sqrt {E2}+ \frac{\hat{q} \cdot \omega }{ \bar{P} \cdot \omega } 
\frac{s_i\!-\!s_f}{2Q}\Big ) }\, ,
\nonumber \\ 
&&f=4\frac {\Big (1 \!+\! 2\frac{\hat{q} \cdot  \omega }{ 
\bar{P} \cdot \omega }\bar{P} \!\cdot\! \hat{q}
\!-\! \Big (\frac{\hat{q} \cdot  \omega }{ \bar{P} \cdot \omega }\Big)^2
\bar{P}^2 \Big)}{ D\Big (\!\sqrt {E2}\!+\!
\frac{\hat{q} \cdot  \omega }{ \bar{P} \cdot \omega } 
\frac{s_i\!-\!s_f}{2Q}
\Big )^2}\, , \;
%
g=1+ \frac{\hat{q} \ccdot \omega }{ \bar{P} \ccdot \omega } \,
\frac{s_i\!-\!s_f}{2Q\sqrt {E2}}\,.\label{eq:gen-ff4}
\end{eqnarray}
The sum symbol, $\sum$, accounts for the fact that there are two acceptable
values of $p^z$ (and $e_p$) corresponding to the same set of variables 
$\bar{s}$, $s_i\!-\!s_f$ and $\frac{p \cdot \omega}{\bar{P} \cdot \omega }$.
Contrary to some cases but similarly to other ones considered in this work, 
the choice of the new variables makes the transformed quantity 
$\frac{d\vec{p}}{e_p}$ be the same for the two values of  $p^z$ 
(and the associated $e_p$).
Inserting the above expression into that one for the charge form factor, 
one gets:
\begin{eqnarray} 
``F_1(Q^2)"&\!=\!&\frac{16\pi^2}{N}\! \int\! \frac{d\vec{p}}{(2\pi)^3}\;
\frac{1}{e_p} \;^{``}\Bigg(
\frac {(p_i\!+\!p_f\!+\!2p)\!\cdot\!\omega}{2(p_i\!+\!p_f)\!\cdot\!\omega}   \;
\tilde{\phi}(\vec{k_f}^2)\;\tilde{\phi}(\vec{k_i}^2) \Bigg)^{"}
\nonumber \\
&\!=\!&\frac{2}{\pi N}\!\int \! \! \int d\bar{s} \; d(\frac{s_i\!-\!s_f}{Q}) 
\frac{  \theta(\cdots)  }{4\sqrt{D}} \;
\phi(s_f)\;\phi(s_i) 
\nonumber \\
&& \hspace*{0.6cm} \times
\sum \int \!\frac {d(\frac{p \cdot \omega}{\bar{P} \cdot \omega })  }{ 
\sqrt {\Big  (\frac {s_i\,s_f}{D }\!-\!m^2\Big )f
-(\frac{p \cdot \omega}{\bar{P} \cdot \omega }\!-\!d)^2}} 
\Bigg (\frac{2\bar{s}\!+\!Q^2 }{ D} 
 -(\frac{p \ccdot \omega}{\bar{P} \ccdot \omega}-d)\,g \Bigg) \, .
\label{eq:gen-ff5} 
\end{eqnarray} 
Performing the last operations in the above equation provides 
a factor $2\pi\frac{2\bar{s}\!+\!Q^2 }{ D} $, 
allowing one to recover  eq. (\ref{eq:ff1-disp2}).

To avoid too many different notations in the main text, we mostly  
expressed the results in terms of the spectator momentum. It is well known 
however that other variables are currently used in the present front-form case 
(rather then referred to as  light-front approach). 
They are the $x$ variable which represents the ``longitudinal" momentum 
($x=\frac{p \cdot \omega}{\bar{P}  \cdot \omega}$) and the component of the
momentum perpendicular to the front orientation, $p_{\perp}$. 
For completeness, we reproduce below some steps in the case 
$q \ccdot \omega=0$, which is generally considered. We believe that showing 
how the dispersion-relation expressions are obtained in this case 
is important. On the one hand, this approach is often considered as more
reliable than other ones. On the other hand, it is rather ignored that the
3-dimensional integration implied by the calculation of form factors for a
two-body system can be reduced to a two-dimensional one.

Without loss of generality, one can assume that the perpendicular components 
of the initial and final momenta have opposite signs. The expressions of 
$\bar{s}$ and $(s_i\!-\!s_f)$ are given by:
\begin{eqnarray} 
\bar{s}=\frac {m^2\!+\!p_{\perp}^2\!+\!x^2\;\frac{Q^2}{4} }{ x\,(1\!-\!x)}, 
\hspace*{0.5cm} 
s_i\!-\!s_f=2\,\frac {\vec{p}_{\perp} \ccdot \vec{Q}_{\perp}  }{ 1\!-\!x}\, ,
\end{eqnarray}
from which we obtain:
\begin{eqnarray}
p_{\perp}^2 =x\,(1\!-\!x)\;\bar{s} \!-\!m^2\!-\!x^2\;\frac{Q^2}{4},
\hspace*{0.5cm} \vec{p}_{\perp} \ccdot \vec{Q}_{\perp}
=(1\!-\!x)\frac{s_i\!-\!s_f}{2}\,.
\end{eqnarray}
These relations can be used to determine the expression of the integration
volume in terms of $\bar{s}$, $(s_i\!-\!s_f)$ and $x$. 
Together with other factors entering the charge form factor, 
this one is given by:
\begin{eqnarray}
\frac{d\vec{p}}{e_p} \; 
\frac {\bar{P} \ccdot \omega}{2(\bar{P}-p) \ccdot \omega}&= &
\frac{d\vec{p}_{\perp}\;dx}{2 x\,(1\!-\!x)}=
\frac{d\bar{s}\; d(s_i\!-\!s_f)\;dx\;x(1\!-\!x)^2
}{ 4 x(1\!-\!x) \, |{\rm sin}\phi_0| \, |\vec{p}_{\perp}| \, Q}
\nonumber \\
&=& \frac {d\bar{s}\; d(s_i\!-\!s_f)\;dx\; (1\!-\!x)}{2\, 
Q\sqrt {4\Big(x\,(1\!-\!x)\;\bar{s}\!-\!m^2\!-\!x^2\;\frac{Q^2}{4} \Big) 
\!-\!(1\!-\!x)^2\;\frac{(s_i\!-\!s_f)^2}{Q^2} }  } 
\nonumber \\
&=& \frac {d\bar{s}\; d(s_i\!-\!s_f)\;dx\; (1\!-\!x) }{ 2\, Q\sqrt {D}
\sqrt {\Big  (\frac {s_i\,s_f}{D }-m^2\Big )f-(x-d)^2}   }\, ,
\label{2p}
\end{eqnarray}
where $\phi$ is the angle between $\vec{p}_{\perp}$ and $\vec{Q}_{\perp}$. 
Notice that a factor $2$ has been introduced at the numerator of the r.h.s.
term at the first line to take 
into account that there are two values of $\phi$ corresponding 
to the same value of ${\rm cos}\phi$. The quantities d and f are given by:
\begin{eqnarray}
d=\frac{2\bar{s}\!+\!\frac{(s_i\!-\!s_f)^2}{Q^2}}{D}\,,\hspace*{0.5cm}
f=\frac{4}{D}\,.
\end{eqnarray}
Plugging the expression of the integration volume in eq. (\ref{eq:gen-ff5}), 
one gets:  
\begin{eqnarray} 
F_1(Q^2) & \!=\! & \frac{2}{\pi N} \int \! \!\int \!d\bar{s} \; 
d(\frac{s_i\!-\!s_f}{Q}) \frac{1}{2\, \sqrt {D}}\;\phi(s_i) \; \phi(s_f)
\times\!\int \!\frac{  dx \;((1\!-\!d)-(x\!-\!d))}{
\sqrt {\Big  (\frac {s_i\,s_f}{D }\!-\!m^2\Big )f-(x\!-\!d)^2}   }
\nonumber \\ 
& = &\frac{1}{N} \int  \! \!\int d\bar{s} \; 
d(\frac{s_i\!-\!s_f}{Q})  \;  
\frac{ \;(2\,\bar{s}+Q^2) \,\theta(\cdots)}{
D^{3/2} } \;\phi(s_i) \; \phi(s_f), 
\label{2q}
\end{eqnarray}
where the integral over $x$ is limited to the positive values of
what is below the square-root factor at the denominator (first line). 
Moreover, it supposes that the coefficient of $f$  in this factor  
be positive, allowing one to recover the condition given 
by eq. (\ref{eq:theta}). The term  $1\!-\!x$ at the numerator 
can be split into two terms proportional to $(1\!-\!d)$ and $(x\!-\!d)$. 
The last one vanishes when integrated over while the other one  provides
the factor $\frac{2\,\bar{s}+Q^2}{D}$. The scalar form factor, $F_0(Q^2)$, 
given below, differs from the charge one by the replacement 
of the factor $1\!-\!x$ by the factor $\frac{1}{2}$. In this simpler 
case, the integration over $x$ provides a factor $\frac{\pi}{2}$  
at the numerator (from $\int \frac{dx}{2(a^2-x^2)^{1/2}}=\frac{\pi}{2}$):
\begin{eqnarray} 
F_0(Q^2) &\! =\! & \frac{2}{\pi N} \int\! \!\int \! d\bar{s} \; 
d(\frac{s_i\!-\!s_f}{Q}) \frac{1}{2\, 
\sqrt {D}} \;\phi(s_i) \; \phi(s_f)
\times \!\int \! 
\frac{ dx }{2\, 
\sqrt {\Big  (\frac {s_i\,s_f}{D }\!-\!m^2\Big )f-(x\!-\!d)^2}   } 
\nonumber \\ 
& = & \frac{1}{N} \int \! \!\int  d\bar{s} \; d(\frac{s_i\!-\!s_f}{Q}) \; 
\frac{ \theta(\cdots)  }{2\, \sqrt{D} } \;\phi(s_i) \; \phi(s_f)\,.
\label{2r}
\end{eqnarray}
%

\subsection{Form factors in the ``point form": arbitrary momentum configuration}
\label{gen:pf}
We start from the expressions of the $s_i,\,s_f$ quantities derived in appendix
\ref{detail:pf} for the elastic case ($\hat{v} \ccdot \hat{q}=0 $). 
The combinations with a symmetry character are given by:
\begin{eqnarray}
&&\bar{s} =2\Bigg( 
\frac{\sqrt{(p \!\cdot\! \hat{v})^2 \!-\! (p \!\cdot\! \hat{q})^2\!+\!\frac{Q^2}{4}}
}{\sqrt{(p \!\cdot\! \hat{v})^2 \!-\! (p \!\cdot\! \hat{q})^2}}\;
\Big((p \!\cdot\! \hat{v})^2 +(p \!\cdot\!\hat{q} )^2\Big)
+(p \!\cdot\! \hat{v})^2 -(p \!\cdot\!\hat{q} )^2 \Bigg) \, ,
\nonumber \\
&&s_i\!-\!s_f= -4\,
\frac{(p \!\cdot\! \hat{v})(p \!\cdot\!q )}{
\sqrt{(p \!\cdot\! \hat{v})^2 \!-\! (p \!\cdot\! \hat{q})^2}}\, .
\label{eq:gen-pf1}
\end{eqnarray}
We first invert the above expressions to express the different quantities 
appearing at the r.h.s. in terms of $s_i, \, s_f$. One gets:
\begin{eqnarray}
p \!\cdot\! \hat{v} = \frac{(s_i\,s_f)^{1/4}\,(\sqrt{s_i}\!+\!\sqrt{s_f})
}{2\sqrt{(\sqrt{s_i}\!+\!\sqrt{s_f})^2\!+\!Q^2}}\, ,\hspace*{1cm}
p \!\cdot\! \hat{q} = -\frac{(s_i\,s_f)^{1/4}\,(\sqrt{s_i}\!-\!\sqrt{s_f})
  }{ 2\sqrt {(\sqrt{s_i}\!-\!\sqrt{s_f})^2\!+\!Q^2} }  \, .
\label{eq:gen-pf2}
\end{eqnarray}
As only the initial and final momenta are involved, it is appropriate to
consider the plane they determine as the $x,\,y$ plane. The  $x,\,y$ components
of $\vec{p}$ can then be determined in terms of $s_i,\, s_f,\, p^z$. They are
more simply expressed using $e_p$:

\begin{eqnarray}
p^x=\frac{(e_p\,\hat{v}^0\!-\!p \!\cdot\! \hat{v})\,\hat{q}^y 
-(e_p\,\hat{q}^0\!-\!p \!\cdot\! \hat{q})\,\hat{v}^y
 }{ \hat{v}^x \,\hat{q}^y - \hat{v}^y \,\hat{q}^x}\, ,\hspace*{0.5cm}
p^y=\frac{(e_p\,\hat{v}^0\!-\!p \!\cdot\! \hat{v})\,\hat{q}^x
-(e_p\,\hat{q}^0\!-\!p \!\cdot\! \hat{q})\,\hat{v}^x
 }{ \hat{v}^y \,\hat{q}^x - \hat{v}^x \,\hat{q}^y}\, ,
\label{eq:gen-pf3}
\end{eqnarray}
with
\begin{eqnarray}
e_p=(\hat{v}^0 p \!\cdot\! \hat{v} \!-\! \hat{q}^0p \!\cdot\! \hat{q})\pm 
\sqrt{\hat{v}^{02} \!-\! \hat{q}^{02} \!-\! 1}\;
\sqrt{(p \!\cdot\! \hat{v})^2\!-\!(p \!\cdot\! \hat{q})^2 \!-m^2\!-\!p^{z2}} \, .
\label{eq:gen-pf4}
\end{eqnarray}
In the change of the variables $p^x,\,p^y,\,p^z$ 
to the variables $\bar{s}$, $s_i\!-\! s_f$, $ p^z$, 
the integration volumes transform as follows: 
\begin{eqnarray}
\frac{d\vec{p}}{e_p}=\sum \frac{1}{4}
\frac{d\bar{s} \; d(s_i\!-\!s_f)\, dp^z}{
Q \sqrt{\frac{s_i\,s_f}{D}
\!-\!m^2\!-\!p^{z2}}} \; \frac{2\bar{s}\!+\! Q^2}{\,D^{3/2}}\,,
\label{eq:gen-pf5}
\end{eqnarray}
where the sum symbol has been introduced to take into account 
that there are two values of $e_p$ to be considered, see eq. (\ref{eq:gen-pf4}). 
A factor $e_p$, which appears in the transformation, has been put at the l.h.s.
so that to recover a standard minimal relativity factor, which, in any case, 
is present in the approach. Inserting 
this relation into that one for the charge form factor, one obtains:
\begin{eqnarray}
&&``F_1(Q^2)"=\frac{16\,\pi^2}{N} \; \int \frac{d \vec{p}}{(2\pi)^3} \; 
\frac{1}{e_p} \;
^{``}\Bigg(\tilde{\phi}(\vec{k}_{f}^2) \;\tilde{\phi}(\vec{k}_{i}^2)\Bigg)^{"}  
\nonumber \\ 
&&\hspace*{1.0cm}=\frac{2}{\pi N} \!\int \! \!  \int\!  
d\bar{s} \; d(\frac{s_i\!-\!s_f}{Q})\;
\frac{(2\bar{s}\!+\! Q^2)\;\theta(\cdots)
}{4 \, D^{3/2}} \;\phi(s_f)\;\phi(s_i) \times\!
\sum \!\int \! \frac{dp^z}{\sqrt{\frac{s_i\,s_f}{D}\!-\!m^2\!-\!p^{z2}}} \,.
\hspace*{0.5cm}
\label{eq:gen-pf6}
\end{eqnarray}
After performing the last operations ($\sum$ and $\int dp^z \cdots$), 
which provides a factor $2\pi$,
eq.~(\ref{eq:ff1-disp2}) is easily recovered. 
As already mentioned, we have only considered here the elastic case for
simplicity. Equations (\ref{eq:gen-pf5}) and (\ref{eq:gen-pf6}) can be shown 
to also hold in the inelastic case while eqs. (\ref{eq:gen-pf1}), 
(\ref{eq:gen-pf2}) and (\ref{eq:gen-pf4}) contain extra 
$\hat{v} \ccdot \hat{q} $ dependent terms.

It is noticed that, instead of showing that the final expression could be
directly cast into the dispersion approach one, the same result could 
be achieved by making a Lorentz transformation from the present general case 
to the Breit frame one. Not surprisingly, the parameters of the transformation 
are determined by the vector $\vec{v} = -\vec{\hat{v}}/\hat{v}^0$.

\subsection{Generalization to an hyperplane with arbitrary $\lambda^{\mu}$}
\label{gen:hyp}
In this subsection, we generalize the instant-form results 
to an arbitrary hyperplane. We first consider the derivation of the factor
$\alpha$ that has to multiply $q$ so that the square momentum transfer 
to the constituents be equal to that one to the system. In the following part, 
we give the main results pertinent to the implementation of this relation for
the charge form factor. The overall size of the 4-vector $\lambda^{\mu}$ being
irrelevant, we use the unit 4-vector defined as 
$\hat{\lambda}^{\mu}=\lambda^{\mu} /\sqrt{\lambda^2} $. 

The square momentum transfer to the struck constituent reads: 
\begin{eqnarray}
&&(p_i\!-\!p_f)^2= q^2\!-\!(q \!\cdot\!\hat{\lambda})^2
+\Bigg ( \sqrt{m^2\!+\!((\bar{P}\!-\!\frac{q}{2}\!-\!p)\!\cdot\!\hat{\lambda})^2
\!-\!(\bar{P}\!-\!\frac{q}{2}\!-\!p)^2} 
\nonumber \\
&&\hspace*{5cm}
-\sqrt{m^2\!+\!((\bar{P}\!+\!\frac{q}{2}\!-\!p)\!\cdot\!\hat{\lambda})^2
\!-\!(\bar{P}\!+\!\frac{q}{2}\!-\!p)^2}
\Bigg)^2 \, .
\end{eqnarray}
It is noticed that the square root factor may be written as:
\begin{eqnarray}
&&\sqrt{m^2\!+\!((\bar{P}\!\pm\!\frac{q}{2}\!-\!p)\!\cdot\!\hat{\lambda})^2 
\!-\!(\bar{P}\!\pm\!\frac{q}{2}\!-\!p)^2}
\nonumber \\
&&\hspace*{1cm}=\sqrt{m^2\!+\!((\bar{P}\!-\!p)\!\cdot\!\hat{\lambda})^2 
   \!-\!(\bar{P}\!-\!p)^2\!-\!\frac{\tilde{q}^2}{4}
\mp \sqrt{-\tilde{q}^2}\, (\bar{P}\!-\!p) \!\cdot\! \hat{\tilde{q}} } \,,
\end{eqnarray}
where, at the last line, we introduced the notation
$\tilde{q}^{\mu}=q^{\mu}- \hat{\lambda}^{\mu} (\hat{\lambda}\!\cdot\! q)$, 
$\hat{\tilde{q}}^{\mu}=\tilde{q}^{\mu}/\sqrt{-\tilde{q}^2}$.
To get the equation to be fulfilled by $\alpha$, 
we assume that the coefficient of the 4-vector 
$\tilde{q}^{\mu}$ 
is multiplied by this quantity, which gives the equation:
\begin{eqnarray}
&&``(p_i\!-\!p_f)^2"= q^2=
\alpha^2\;\tilde{q}^2 
\nonumber \\
&&\hspace*{1cm}+\Bigg ( \sqrt{m^2\!+\!((\bar{P}\!-\!p)\!\cdot\!\hat{\lambda})^2 
\!-\!(\bar{P}\!-\!p)^2\!-\!
\frac{\alpha^2\;\tilde{q}^2}{4}
+ \alpha \sqrt{-\tilde{q}^2}\, (\bar{P}\!-\!p) \!\cdot\! \hat{\tilde{q}}} 
\nonumber \\
&&\hspace*{2cm}
-\sqrt{m^2\!+\!((\bar{P}\!-\!p)\!\cdot\!\hat{\lambda})^2 
\!-\!(\bar{P}\!-\!p)^2\!-\!
\frac{\alpha^2\;\tilde{q}^2}{4}
- \alpha \sqrt{-\tilde{q}^2}\, (\bar{P}\!-\!p) \!\cdot\! \hat{\tilde{q}}}
\Bigg)^2 \, .
\end{eqnarray}
Its solution is given by:
\begin{eqnarray}
\alpha^2=\frac{q^2}{\tilde{q}^2}\;
\frac{m^2+((\bar{P}\!-\!p)\!\cdot\! \hat{\lambda})^2 -(\bar{P}\!-\!p)^2\!-\!\frac{q^2}{4}
}{m^2+((\bar{P}\!-\!p)\!\cdot\! \hat{\lambda})^2 -(\bar{P}\!-\!p)^2
- \Big( (\bar{P}\!-\!p) \!\cdot\! \hat{\tilde{q}} \Big)^2\! -\!\frac{q^2}{4} }
\,.
\end{eqnarray}
It represents some relatively straightforward generalization 
of eq. (\ref{alpif}).\\
An interesting and very useful relation which follows from implementing  
the constraints related to  space-time translation invariance is given by:
\begin{eqnarray}
&&``\;\sqrt{m^2\!+\!((\bar{P}\!\pm\!\frac{q}{2}\!-\!p)\!\cdot\! \hat{\lambda})^2 
\!-\!(\bar{P}\!\pm\!\frac{q}{2}\!-\!p)^2} \;"
=\sqrt{m^2\!+\!((\bar{P}\!\pm\!\frac{\alpha q}{2}\!-\!p)\!\cdot\! \hat{\lambda})^2 
\!-\!(\bar{P}\!\pm\!\frac{\alpha q}{2}\!-\!p)^2}
\nonumber \\
&&\hspace*{1cm}=\sqrt{m^2\!+\!((\bar{P}\!-\!p)\!\cdot\! \hat{\lambda})^2 
\!-\!(\bar{P}\!-\!p)^2\!-\!\frac{q^2}{4}}
\nonumber \\
&&\hspace*{1.5cm}\mp\frac{1}{2}\sqrt{\frac{q^2}{\tilde{q}^2}}\;
\frac{(\bar{P}\!-\!p) \!\cdot\! \tilde{q} }{
\sqrt{m^2\!+\!((\bar{P}\!-\!p)\!\cdot\! \hat{\lambda})^2 \!-\!(\bar{P}\!-\!p)^2
\!-\!\Big( (\bar{P}\!-\!p) \!\cdot\! \hat{\tilde{q}} \Big)^2
    \!-\!\frac{q^2}{4}  } } \,.
\end{eqnarray}
The quantities $s_i,\; s_f$ can be expressed as:
\begin{eqnarray}
s_{i,f}&\!=\!& 2(p\!\cdot\! \hat{\lambda})^2\!+\!2\bar{P}\!\cdot\!p
\!-\!2\bar{P}\!\cdot\! \hat{\lambda} \, p\!\cdot\! \hat{\lambda}
\!+\!2p\!\cdot\! \hat{\lambda} \,
\sqrt{m^2\!+\!((\bar{P}\!-\!p)\!\cdot\! \hat{\lambda})^2 
\!-\!(\bar{P}\!-\!p)^2\!-\!\frac{q^2}{4}}
\nonumber \\
&&\pm\sqrt{-q^2} \; 
\frac {  p\!\cdot\! \hat{\lambda}\;(\bar{P}\!-\!p) \!\cdot\! \hat{\tilde{q}} 
 -p \!\cdot\! \hat{\tilde{q}}\,
 \sqrt{m^2\!+\!((\bar{P}\!-\!p)\!\cdot\! \hat{\lambda})^2 
\!-\!(\bar{P}\!-\!p)^2\!-\!\frac{q^2}{4}}
}{\sqrt{m^2\!+\!((\bar{P}\!-\!p)\!\cdot\! \hat{\lambda})^2 \!-\!(\bar{P}\!-\!p)^2
\!-\!\Big( (\bar{P}\!-\!p) \!\cdot\! \hat{\tilde{q}} \Big)^2
    \!-\!\frac{q^2}{4}  }}\,.
\end{eqnarray}

We now consider the calculation of the charge form factor. In
order to somewhat simplifying the writing of the expressions, we introduce 
the notation  
$\tilde{\bar{P}}^{\mu}= \bar{P}^{\mu}-\hat{\lambda}^{\mu} \, (\bar{P}\!\cdot\! \hat{\lambda}) $.
The combinations of  $s_i,\; s_f$ with a symmetry character are now given by:
\begin{eqnarray}
&&\bar{s}= 2(p\!\cdot\! \hat{\lambda})^2\!+\!2p\!\cdot\!\tilde{\bar{P}}
\!+\!2p\!\cdot\! \hat{\lambda} \,
\sqrt{(p\!\cdot\! \hat{\lambda})^2\!-\!\tilde{\bar{P}}^2
\!+\!2p\!\cdot\!\tilde{\bar{P}} \!+\!\frac{Q^2}{4}}\, ,
\nonumber \\
&&s_i\!-\!s_f= 2Q \; 
\frac {p\!\cdot\! \hat{\lambda}\,(\bar{P}\!-\!p) \!\cdot\! \hat{\tilde{q}}
-p \!\cdot\! \hat{\tilde{q}}\,
\sqrt{(p\!\cdot\! \hat{\lambda})^2\!-\!\tilde{\bar{P}}^2
\!+\!2p\!\cdot\!\tilde{\bar{P}} \!+\!\frac{Q^2}{4}}
}{\sqrt{
(p\!\cdot\! \hat{\lambda})^2\!-\!\tilde{\bar{P}}^2
\!+\!2p\!\cdot\!\tilde{\bar{P}}
\!-\!\Big( (\bar{P}\!-\!p) \!\cdot\! \hat{\tilde{q}} \Big)^2  \!+\!\frac{Q^2}{4} 
 }}\, .
\end{eqnarray}
The above expressions can be inverted to get the quantities 
$p\!\cdot\!\tilde{\bar{P}}$ and $p \!\cdot\! \hat{\tilde{q}}$  
in terms of the variables  $\bar{s}$, $s_i- s_f$ and 
$p\!\cdot\! \hat{\lambda}$, 
with the result:
\begin{eqnarray}
&&(p \!\cdot\!\tilde{\bar{P}})=
\frac{1}{2} \Big(\bar{s}-p\!\cdot\! \hat{\lambda}\,\sqrt{D0} \Big)\, ,
\nonumber \\
&&(p \!\cdot\!\hat{\tilde{q}})=
 p\!\cdot\! \hat{\lambda}\,
 \frac{2 \bar{P} \!\cdot\!\hat{\tilde{q}} \sqrt{D0}
 \!+\!\frac{s_i\!-\!s_f}{Q}\sqrt{D2} }{D1}
- \frac{s_i\!-\!s_f}{2Q} \, \frac{\sqrt{D0\;D2} \!-\!
2 \bar{P} \!\cdot\!\hat{\tilde{q}}\,\frac{s_i\!-\!s_f}{Q}}{D1}
 \, , \label{eq:gen-hyp8}
\end{eqnarray}
where the $D$ quantities generalize those given in appendix \ref{gen:if}:
\begin{eqnarray}
&&D0=4\bar{s} \!+\! Q^2 \!-\! 4\tilde{\bar{P}}^2\, ,
\nonumber \\
&&D1= 4\bar{s} \!+\! Q^2\!+\! \frac{(s_i\!-\!s_f)^2}{Q^2}
  \!-\! 4\tilde{\bar{P}}^2\, ,
\nonumber \\
&&D2= 4\bar{s} \!+\! Q^2\!+\! \frac{(s_i\!-\!s_f)^2}{Q^2}
  \!-\! 4\tilde{\bar{P}}^2\!-\!4(\bar{P} \!\cdot\!\hat{\tilde{q}})^2\, .
\end{eqnarray}
The Jacobian of the transformation can be calculated in two steps,  
$p^x,\,p^y,\,p^z$ to $p\!\cdot\!\tilde{\bar{P}}$ and $p \!\cdot\! \hat{\tilde{q}}$ 
and $p\!\cdot\! \hat{\lambda}$  and then from 
$p\!\cdot\!\tilde{\bar{P}}$ and $p \!\cdot\! \hat{\tilde{q}}$ 
and $p\!\cdot\! \hat{\lambda}$ to $\bar{s}$, $s_i\!-\!s_f$ 
and $p\!\cdot\! \hat{\lambda}$.
The first one is rather cumbersome while the second one is rather easy. 
For the first one, the result given in appendix \ref{gen:jac}, could be useful.
The resulting expression for the transformation of the integration volume 
is given by:
\begin{eqnarray}
&&\frac{d\vec{p}}{e_p}\,\frac{``(2p\!+\!p_i\!+\!p_f)"\!\cdot\! \lambda 
}{2``(p_i\!+\!p_f)" \!\cdot\! \lambda} =
\sum \frac {d\bar{s} \; d(s_i\!-\!s_f)\,d(p\!\cdot\! \hat{\lambda}) \;
\Big((2\bar{s} \!+\! Q^2)- (p\!\cdot\! \hat{\lambda}\!-\!d)g\Big)
}{4Q\,D\sqrt{D} \sqrt {\Big (\frac{s_is_f}{D}\!-\!m^2  \Big )f \!-\! 
(p\!\cdot\! \hat{\lambda}\!-\!d)^2}} \, ,\label{eq:gen-hyp10}
\end{eqnarray}
where $D$ has already been defined while:
\begin{eqnarray}
&&d=\frac{\sqrt{D2}\Big (\bar{s}-2\tilde{\bar{P}}^2\,
\frac{2\bar{s}\!+\! \frac{(s_i\!-\!s_f)^2}{Q^2}}{D}\Big)
-\bar{P}\!\cdot\!\hat{\tilde{q}}\,\frac{s_i\!-\!s_f}{Q}\sqrt{D0}
}{\Big (\sqrt{D0\;D2} \!-\!2
\bar{P}\!\cdot\!\hat{\tilde{q}}\,\frac{s_i\!-\!s_f}{Q}\Big)}  \,,
\nonumber \\
&&f=
\frac{4D1^2( -\tilde{\bar{P}}^2\! -   \!  (\tilde{\bar{P}} \!\cdot\!\hat{\tilde{q}})^2 ) 
 }{D\Big (\sqrt{D0\;D2} \!-\!2
\bar{P}\!\cdot\!\hat{\tilde{q}}\,\frac{s_i\!-\!s_f}{Q}\Big)^2 }\, ,\hspace*{0.5cm}
g=2\frac {D \Big (\sqrt{D0\;D2} \!-\!2
\bar{P}\!\cdot\!\hat{\tilde{q}}\,\frac{s_i\!-\!s_f}{Q}\Big)}{D1\sqrt{D2}} \,.
\end{eqnarray}
Like in other cases, the sum symbol takes into account that there are two
values of $p^z$ (and $e_p$) to be considered. Similarly to the front-form case,
appendix \ref{gen:ff}, the transformed expression of $\frac{d\vec{p}}{e_p}$ is
the same for the two values.
The above expression can now be inserted in that one for the form factor:
\begin{eqnarray}
&&\hspace*{-0.6cm}``F_1(Q^2)"=\frac{16\pi^2}{N(2\pi)^3}\!
\int \frac{d\vec{p}}{e_p}\,\frac{``(2p\!+\!p_i\!+\!p_f)"\!\cdot\! \lambda 
}{2``(p_i\!+\!p_f)" \!\cdot\! \lambda}\;
^{``}\Bigg(\tilde{\phi}(\vec{k_f}^2)\;\tilde{\phi}(\vec{k_i}^2)\Bigg)^{"} 
\nonumber \\
&&\hspace*{-0.3cm}=\!\frac{2}{\pi N}\!
\int \! \!\int \! d\bar{s} \; d(\frac{s_i\!-\!s_f}{Q})\;
\frac{\theta(\cdots) }{4\,D^{3/2}}\; \phi(s_f) \; \phi(s_i)
\!\times \!\sum \!\int \!\frac { d(p \ccdot \hat{\lambda})
\Big ((2\bar{s}\!+\! Q^2)-(p\!\cdot\! \hat{\lambda}\!-\!d)g\Big)
}{\sqrt{\Big( \frac{s_is_f}{D} \!-\!m^2\Big)f \!-\! 
(p\!\cdot\! \hat{\lambda}\!-\!d)^2}}\,.\hspace{0.6cm}\label{eq:gen-hyp12}
\end{eqnarray}
Noting that the last operations provide a factor $2\pi (2\bar{s}\!+\! Q^2)$,
 eq. (\ref{eq:ff1-disp2}) is recovered. 

Instead of showing a direct relation of the above results to the
dispersion-relation ones, we could as well make a Lorentz transformation 
with parameter $\vec{v}=-\vec{\hat{\lambda}}/\hat{\lambda}^0$. 
This brings the present results back to the instant-form ones 
(appendices \ref{detail:if}, \ref{gen:if}), 
taking into account that quantities such as $\bar{P}^{\mu}$ or $q^{\mu}$ 
should be changed in the transformation. 
An other remark concerns the relation of results 
in this section to the front-form ones, appendix \ref{gen:ff}. 
In principle, it is expected that the latter could be obtained 
from the former in the limit $\lambda^2 \rightarrow 0$. 
Taking this limit is not straightforward however and requires some care. 
The two sets of results, derived independently,  
appear to verify the expected relation.

\subsection{Another solution}
\label{gen:other}
As mentioned in the text, there  may be many solutions to restore properties
related to space-time translations. We here present another one for the
front-form case in the Breit frame and a momentum transfer perpendicular 
to the front orientation. It is obtained by assuming that $\bar{E}$ 
is not an independent quantity and could be affected by the change 
$q^{\mu} \rightarrow \alpha q^{\mu}$ as follows 
$\bar{E}=\sqrt{M^2-\frac{q^2}{4}} \rightarrow
\sqrt{M^2-\frac{\alpha^2q^2}{4}}$.
Using the components of the spectator momentum, $\vec{p}$, parallel and
perpendicular to the front orientation, $p_{\parallel}$ and $p_{\perp}$, the
new factor $\alpha'\,^2$ reads:
\begin{eqnarray}
&&\alpha'\,^2=M^2\Bigg (\frac{  \sqrt{m^2_{\perp}+\frac{Q^2}{4}} - 
\sqrt{m^2_{\perp}+\frac{Q^2}{4}(\frac{e_p+p_{\parallel}}{M})^2} \,(\frac{e_p+p_{\parallel}}{M})
}{\sqrt{m^2_{\perp}+\frac{Q^2}{4}} \sqrt{m^2_{\perp}+\frac{Q^2}{4}\,(\frac{e_p+p_{\parallel}}{M})^2}
 +\frac{Q^2}{4}(\frac{e_p+p_{\parallel}}{M})}\Bigg)^2\, ,
\nonumber \\
&&{\rm in \;place\; of} \hspace{1cm} \alpha^2= 
\frac{(\sqrt{M^2+\frac{Q^2}{4}}-(e_p+p_{\parallel}) )^2
}{m^2_{\perp}+\frac{Q^2}{4}} \, ,
\end{eqnarray}
where $m^2_{\perp}=m^2+p^2_{\perp}$. The corresponding expression 
of the charge form factor reads:
\begin{eqnarray} 
&& \hspace*{-1.0cm}``F_1(Q^2)"\!=\!\frac{16\pi^2}{N}\! \int\! \frac{d\vec{p}}{(2\pi)^3}\;
\frac{1}{e_p} \;
\frac{\sqrt{m^2_{\perp}+\frac{Q^2}{4}} \sqrt{m^2_{\perp}+\frac{Q^2}{4}\,(\frac{e_p+p_{\parallel}}{M})^2}
 +\frac{Q^2}{4}(\frac{e_p+p_{\parallel}}{M})}{  \sqrt{m^2_{\perp}+\frac{Q^2}{4}} - 
\sqrt{m^2_{\perp}+\frac{Q^2}{4}(\frac{e_p+p_{\parallel}}{M})^2} \,\frac{e_p+p_{\parallel}}{M}}
\nonumber \\
&&  \hspace{5 cm}  \times
\frac{1}{2\sqrt{m^2_{\perp}+\frac{Q^2}{4}(\frac{e_p+p_{\parallel}}{M})^2}}
^{``}\Bigg(\tilde{\phi}(\vec{k_f}^2)\;\tilde{\phi}(\vec{k_i}^2) \Bigg)^{"}\,,
\nonumber \\
&& \hspace*{-1.0cm} {\rm in \;place\; of} \hspace{0.3cm} 
``F_1(Q^2)"\!=\!\frac{16\pi^2}{N}\! \int\! \frac{d\vec{p}}{(2\pi)^3}\,
\frac{1}{e_p} \,
\frac {\sqrt{M^2+\frac{Q^2}{4}}
}{2 \Big(\sqrt{M^2\!+\!\frac{Q^2}{4}}\!-\!(e_p\!+\!p_{\parallel})\Big) } 
^{``}\Bigg(\!\tilde{\phi}(\vec{k_f}^2)\,\tilde{\phi}(\vec{k_i}^2) \Bigg)^{"}
\hspace*{-2mm}.\label{eq:ffother}
\end{eqnarray} 
Not surprisingly, results for the solution considered in this subsection
coincide with those given in the main text at $Q^2=0$.
\subsection{Jacobian}
\label{gen:jac}
In most calculations presented in this work, an important step is the
determination of the Jacobian relative to the transformation 
of the $\vec{p}$ variables
to the set $\bar{s}$, $s_i-s_f$ and a third one. In a few cases, 
this determination can be done in two steps, from variables $\vec{p}$ 
to Lorentz-like scalar ones represented by quantities $p\!\cdot\! a$, 
$p\!\cdot\! b$ and $p\!\cdot\! c$ ($a$, $b$  and $c$ represent 4-vectors),
and from these last variables to $\bar{s}$, $s_i\!-\!s_f$ and the third one. 
The first step, which is generally the most complicated and provides 
the factor $\frac{1}{(a^2-z^2)^{1/2}}$ that will be integrated over 
the third variable in most cases, is given by: 
\begin{eqnarray}
\frac{d\vec{p}}{e_p}= |J_1| \;
d(p\!\cdot\! a)\, d(p\!\cdot\! b) \,d(p\!\cdot\! c)\, ,\label{eq:jac1}
\end{eqnarray}
where $|J_1|$ can be written in terms of Lorentz invariant quantities 
such as $p^2$, $a^2$, $b^2$, $c^2$ and the mixed scalar products 
$p \!\cdot\! a$, $\cdots$ :
\begin{eqnarray}
&&|J_1|=|p^2 a^2 b^2 c^2
\nonumber \\
&&\hspace*{1cm}\!-\! p^2 a^2(b\!\cdot\! c)^2 \!-\! p^2 b^2(a\!\cdot\! c)^2
\!-\! p^2 c^2(a\!\cdot\! b)^2 \!-\! a^2 b^2(p\!\cdot\! c)^2 
\!-\! a^2 c^2(p\!\cdot\! b)^2 \!-\! b^2 c^2(p\!\cdot\! a)^2
\nonumber \\
&&\hspace*{1cm}\!+\!2 p^2 (a\!\cdot\! b)(a\!\cdot\! c)(b\!\cdot\! c) 
\!+\!2 a^2(p\!\cdot\! b)(p\!\cdot\! c)(b\!\cdot\! c)
\!+\!2 b^2(p\!\cdot\! a)(p\!\cdot\! c)(a\!\cdot\! c)
\!+\!2 c^2(p\!\cdot\! a)(p\!\cdot\! b)(a\!\cdot\! b)
\nonumber \\
&&\hspace*{1cm}\!+\! (p\!\cdot\! a)^2 (b\!\cdot\! c)^2\!+\! (p\!\cdot\! b)^2 (a\!\cdot\! c)^2
\!+\! (p\!\cdot\! c)^2 (a\!\cdot\! b)^2
\nonumber \\
&&\hspace*{1cm}\!-\!2(p\!\cdot\! a) (p\!\cdot\! b)(a\!\cdot\! c)(b\!\cdot\! c)
\!-\!2(p\!\cdot\! a) (p\!\cdot\! c)(a\!\cdot\! b)(b\!\cdot\! c)
\!-\!2(p\!\cdot\! b) (p\!\cdot\! c)(a\!\cdot\! b)(a\!\cdot\! c)|^{-\frac{1}{2}}
\,.\label{eq:jac2}
\end{eqnarray}
In case the 4-vectors $b$ and $c$ are orthogonal to the 4-vector $a$, 
the expression greatly simplifies to read:
\begin{eqnarray}
|J_1|=|(p^2 a^2\!-\! (p\!\cdot\! a)^2)( b^2 c^2\! -\!(b\!\cdot\! c)^2) 
\!+\!2a^2 (p\!\cdot\! b)(p\!\cdot\! c)(b\!\cdot\! c)
\!-\! a^2 c^2 (p\!\cdot\! b)^2 
\!-\! a^2 b^2(p\!\cdot\! c)^2|^{-\frac{1}{2}}  .\label{eq:jac3}
\end{eqnarray}
This result can be applied to various cases considered in this work.\\ 
- Front form in the Breit frame, sect. \ref{ssec:ff}
($a^{0,x,y,z}=0,0,0,1$; $b^{0,x,y,z}=0,\hat{n}^x,\hat{n}^y,0$; 
 $c^{0,x,y,z}=0,\hat{q}^x,\hat{q}^y,0$):
\begin{eqnarray}
|J_1|=|(m^2 \!+\! p^{z2})
(-1\!+\! (\hat{n} \ccdot \hat{q})^2)
+2 \vec{p} \ccdot \hat{q} \;\vec{p} \ccdot \hat{n} \;
\hat{n} \ccdot \hat{q}
-(\vec{p} \ccdot \hat{n} )^2 
-(\vec{p} \ccdot \hat{q})^2|^{-\frac{1}{2}}.
\end{eqnarray}
Accounting for eqs. (\ref{eq:rqm-ff2}) together with eq. (\ref{eq:rqm-ff3}) 
allows one to recover, up to some factor, the denominator appearing at the
r.h.s. of the first line in eq. (\ref{eq:rqm-ff4}). \\
- Function $I(s_i,Q^2,s_f)$, appendix \ref{app:fnI}
($a^{0,x,y,z}=0,0,0,1$; $b^{\mu}=\tilde{P}_i^{\mu}$; 
 $c^{\mu}=\tilde{P}_f^{\mu}$):
\begin{eqnarray}
|J_1|=|(m^2 \!+\! p^{z2})
((\tilde{P}_i \ccdot \tilde{P}_f)^2\!-\!\tilde{P}_i^2 \tilde{P}_f^2 ) 
\!-\!2 p \ccdot \tilde{P}_i  \;p \ccdot \tilde{P}_f \;
\tilde{P}_i \ccdot \tilde{P}_f
\!+\!\tilde{P}_i^2  (p \ccdot \tilde{P}_f )^2
\!+\!\tilde{P}_f^2  (p \ccdot \tilde{P}_i )^2|^{-\frac{1}{2}}\,.
\end{eqnarray}
After replacing $\tilde{P}_{i,f}^{\mu}$ in terms 
of $p_i^{\mu}$, $p_f^{\mu}$,  $p^{\mu}$, one finds, 
up to a factor, the denominator in eq. (\ref{eq:fnI4}). 
Contrary to other cases, getting the factor $D$ under the square-root symbol 
at the denominator (or $\sqrt{D}$ in eq. (\ref{eq:fnI5})) 
is quite easy in the present one (see eq. (\ref{eq:sbar})). \\
- Instant form with arbitrary momentum configuration, appendix \ref{gen:if}
 ($a^{0,x,y,z}=0,0,0,1$; $b^{0,x,y,z}=0,\bar{P}^x,\bar{P}^y,0$; 
 $c^{0,x,y,z}=0,\hat{q}^x,\hat{q}^y,0$):
\begin{eqnarray}
|J_1|=|(m^2 \!+\! p^{z2})
(-\vec{\bar{P}}^2\!+\! (\vec{\bar{P}} \ccdot \hat{q})^2)
+2 \vec{p} \ccdot \hat{q} \;\vec{p} \ccdot \vec{\bar{P}} \;
\vec{\bar{P}}\ccdot \hat{q}
-(\vec{p} \ccdot \vec{\bar{P}})^2 
-\vec{\bar{P}}^2(\vec{p} \ccdot \hat{q})^2|^{-\frac{1}{2}}.
\end{eqnarray}
Using eqs. (\ref{eq:gen-if2}) together with eq. (\ref{eq:gen-if4}), 
one recovers, up to a factor, the denominator 
appearing in eq. (\ref{eq:gen-if5}).\\ 
- ``Point form" with arbitrary momentum configuration, appendix \ref{gen:pf} 
($a^{0,x,y,z}=0,0,0,1$; $b^{\mu}=\hat{v}^{\mu}$; $c^{\mu}=\hat{q}^{\mu}$):
\begin{eqnarray}
|J_1|=|(m^2 \!+\! p^{z2})(1\!+\! (\hat{v}\ccdot \hat{q})^2)
-2 p \ccdot \hat{q} \;p \ccdot \hat{v} \;\hat{v} \ccdot \hat{q}
-(p \ccdot \hat{v})^2 +(p \ccdot \hat{q})^2|^{-\frac{1}{2}}\,.
\end{eqnarray}
Using eqs. (\ref{eq:gen-pf2}) together with the limit $\hat{v}\ccdot \hat{q}=0$, 
one recovers, up to a factor, the denominator 
appearing in eq. (\ref{eq:gen-pf5}).\\ 
- Generalized hyperplane, appendix \ref{gen:hyp} 
($a=\hat{\lambda}$; $b=\tilde{\bar{P}}$; $c=\hat{\tilde{q}}$):
\begin{eqnarray}
|J_1|=|(m^2 \!-\! (p\!\cdot\! \hat{\lambda})^2)
( \tilde{\bar{P}}^2\! +   \!  (\tilde{\bar{P}} \!\cdot\!\hat{\tilde{q}})^2 ) 
\!-\!2 (p \!\cdot\!\hat{\tilde{q}}) (p \!\cdot\!\tilde{\bar{P}}) 
(\tilde{\bar{P}} \!\cdot\!\hat{\tilde{q}})
\! -\!(p \!\cdot\!\tilde{\bar{P}})^2
\!+\!\tilde{\bar{P}}^2  (p \!\cdot\!\hat{\tilde{q}})^2 |^{-\frac{1}{2}}\,.
\end{eqnarray}
Using eqs. (\ref{eq:gen-hyp8}), one recovers, up to a factor, the denominator 
appearing in eq. (\ref{eq:gen-hyp10}). \\
Equation (\ref{eq:jac2}) can also be used for the front-form case 
with arbitrary momentum confi\-gu\-ration, appendix \ref{gen:ff}, assuming 
$a=\omega$, $b=\bar{P}$ and $c=\hat{q}$. 
Not much simplification occurs in this case apart from the fact 
that the 4-vector, $a=\omega$  verifies the equality $a^2=0$. Using the
corresponding expression, together with eqs. (\ref{eq:gen-ff2}), 
allows one to recover, up to some factor, the denominator 
at the second line of eq. (\ref{eq:gen-ff3}). 



\begin{thebibliography}{99}

\bibitem{Dirac:1949cp}
P.A.M. Dirac, Rev. Mod. Phys.  21, 392 (1949).

\bibitem{Keister:sb}
B. Keister, W. Polyzou, Adv. Nucl. Phys. 20, 225 (1991). 

\bibitem{Amghar:2002jx} 
A. Amghar, B. Desplanques,   L. Theu{\ss}l, 
Nucl. Phys.  A 714, 213 (2003). 

\bibitem{Junhe:2004}  Jun He, B. Julia-Diaz, Yu-bing Dong, 
Phys. Lett. B 602, 212 (2004).

\bibitem{Julia-diaz:2004}
B. Julia-Diaz, D.O. Riska, F. Coester, Phys. Rev. C 69, 035212 (2004), 
{\it Erratum} Phys. Rev. C 75, 069902 (2007).

\bibitem{Plessas:2004}
W. Plessas, in Proceedings of the Workshop of the 
{\it Physics of Excited Nucleons}, 
Grenoble (France), March 24--27, 2004 (NSTAR 2004),
eds. J.P. Bocquet {\it et al.} (World Scientific).

\bibitem{Allen:2000ge}
T.W. Allen, W.H. Klink, W.N. Polyzou, 
Phys. Rev. C 63, 034002 (2001).

\bibitem{Chung:1988}
P.L. Chung {\it et al.}, Phys. Rev. C 37, 2000 (1988).

\bibitem{Vanorden:1995}
J.W. Van Orden, N. Devine, F. Gross, Phys. Rev. Lett. 75, 4369 (1995).

\bibitem{Desplanques:2004sp}  
B. Desplanques, nucl-th/0407074.

\bibitem{Anisovich:1992}
V.V. Anisovich {\it et al.}, Nucl. Phys.  A 544, 747 (1992). 

\bibitem{Krutov:2002}
A.F. Krutov,  V.E. Troitsky, Phys. Rev. C 65, 045501 (2002).

\bibitem{Melikhov} 
D. Melikhov, hep-ph/0110087, 
Eur. Phys. J. direct C4, 2 (2002).

\bibitem{Wick:1954}
 G.C. Wick, Phys. Rev.  96, 1124 (1954).  

\bibitem{Cutkosky:1954}
 R.E. Cutkosky, Phys. Rev.  96, 1135 (1954).

\bibitem{Lev:1993}
 F.M. Lev, Rivista del Nuovo Cimento 16, 1 (1993). 
 
\bibitem{Bakamjian:1953kh} 
B. Bakamjian, L.H. Thomas, Phys. Rev. 92, 1300 (1953).

\bibitem{Bakamjian:1961} 
B. Bakamjian, Phys. Rev. 121, 1849 (1961).

\bibitem{Sokolov:1985jv}
S.N. Sokolov,  Theor. Math. Phys. 62, 140 (1985).

\bibitem{Desplanques:2004} 
B. Desplanques,  L. Theu{\ss}l, Eur. Phys. J. A 21, 93 (2004).

\bibitem{Coester:2003zh}
F. Coester, Few-Body Suppl. 15, 219 (2002).

\bibitem{Desplanques:2004rd}
B. Desplanques, Nucl. Phys. A 748, 139 (2005).

\bibitem{Wagenbrunn:2001}
R.F. Wagenbrunn {\it et al.}, Phys. Lett. B 511, 33 (2001). 

\bibitem{Sokolov:1978}
S.N. Sokolov, A.N. Shatnii, Theor. Math. Phys. 37, 1029 (1978).




\end{thebibliography}
\end{document}